\documentclass[aps,pre,onecolumn, 10.5pt]{revtex4}
\usepackage{graphicx}
\usepackage{epsfig}
\usepackage{amsmath,amsfonts,enumerate,color}
\usepackage{bm}
\newcommand{\pa}{\partial}

\newcommand{\B}[1]{{\bm{#1}}}
\newcommand{\C}[1]{{\mathcal{#1}}}

\begin{document}

\title{Recent developments in dynamic fracture:\\ Some perspectives
}


\author{Jay Fineberg$^1$ and Eran Bouchbinder$^2$}

\affiliation{$^1$Racah Institute of Physics, Hebrew University of Jerusalem, Jerusalem 91904, Israel\\
$^2$Chemical Physics Department, Weizmann Institute of Science, Rehovot 7610001, Israel}

\begin{abstract}
We briefly review a number of important recent experimental and theoretical developments in the field of dynamic fracture. Topics include experimental validation of the equations of motion for straight tensile cracks (in both infinite media and strip geometries), validation of a new theoretical description of the near-tip fields of dynamic cracks incorporating weak elastic nonlinearities, a new understanding of dynamic instabilities of tensile cracks in both 2D and 3D, crack front dynamics, and the relation between frictional motion and dynamic shear cracks. Related future research directions are briefly discussed.

\end{abstract}

\maketitle

\section{Introduction}
\label{intro}

The goal of this paper is to offer some perspectives on a number of significant developments in the field of dynamic fracture that have taken place over the past decade or so. The subjects that we chose to describe are topics that we have worked on, so this paper is, by definition, quite subjective. We believe that in these chosen areas there has been significant progress --- and each is interesting in its own right. We will make no attempt to present any of these subjects in depth. The idea is to present a schematic, non-technical, overview of the main discoveries and fundamental insights that have evolved from recent research in each of these different areas. Many of these topics have been driven forward by new experiments that have utilized a variety of new techniques to obtain measurements that had not previously been possible. In these cases, comparison of these experiments to the existing theoretical predictions revealed previously unappreciated gaps in our understanding. These, in turn, provided incentives to expand the theoretical framework to encompass these observations. Some, but not all, of the topics addressed in this perspective piece were systematically reviewed in two relatively recent review papers \cite{Bouchbinder.14,BFM10}. Our goal here, as explained above, is different.

The structure of this paper is as follows. In Sect.~\ref{sec:eom} we consider the equation of motion for dynamic tensile (mode I) cracks that has been predicted in the framework of Linear Elastic Fracture Mechanics (LEFM). We will describe recent experiments that have quantitatively shown that the ideas of energy balance that had been the cornerstone of dynamic fracture mechanics indeed provide an excellent quantitative description of crack dynamics as long as rapid fracture is mediated by single ``simple'' cracks; i.e. cracks that propagate along prescribed paths and have not undergone dynamic instabilities.

Section~\ref{sec:wnfm} considers the form of the singular fields surrounding the tips of these ``simple'' cracks. We first discuss a few rather peculiar features of the $\sqrt{r}$ singular fields predicted by LEFM. We then show that experiments which utilize soft materials to ``slow down" crack dynamics, making them accessible to direct measurements, revealed that the  fields in the vicinity of dynamic crack tips may significantly deviate from the LEFM predictions. These fields were explained by theoretically incorporating the leading nonlinear elastic behavior of the medium that becomes important at the high strains that invalidate LEFM as the tip is approached. This theory also predicts the existence of a new intrinsic length scale emerging from the competition between linear and nonlinear elastic constitutive behaviors. Thus, while energy balance predicts the overall dynamics of simple cracks, the structure of the elastic fields driving fracture is modified by elastic nonlinearities near the crack's tip, where linear elasticity starts to break down.

In Sect.~\ref{sec:Instabilities} we go on to consider tensile cracks that are no longer simple. This happens when simple cracks become unstable to different instabilities. We briefly mention the micro-branching instability, where frustrated microscopic branches are spawned at the crack tip, highlighting both its role in significantly limiting crack velocities and its 3D nature. We then show that when micro-branching is suppressed by approaching the 2D limit, cracks accelerate to extremely high velocities, where they become unstable to spontaneous shape oscillations at extreme velocities. Recent theoretical and experimental developments have shed new light on the origin of this instability, and its relation to crack tip nonlinearities through an intrinsic length scale.

The results discussed in Sects.~\ref{sec:eom}-\ref{sec:Instabilities} mainly focus on cracks that propagate in an effectively 2D medium. Section~\ref{sec:3D} discusses phenomena that entail an intrinsically 3D description of fracture dynamics. Based on very recent experiments, we provide additional insight into the properties and origin of the micro-branching instability and provide evidence that it is related to the oscillatory instability. We then briefly discuss both in-plane and out-of-plane crack front dynamics.

While Sects.~\ref{sec:eom}-\ref{sec:3D} consider cracks that are ostensibly tensile in nature, Sect.~\ref{sec:friction} briefly reviews studies that reveal relations between frictional motion and dynamic shear (mode II) cracks.  Experiments over the last decade or so have revealed tight connections between interfacial friction dynamics and crack-like ruptures that appear similar to earthquakes along natural faults. Recent experiments, moreover, suggest that at least some of these can be described with classical mode II crack solutions. Understanding others requires richer coupling of bulk elasticity to constitutive descriptions of friction.

We conclude with some opinions on various directions in which the field of dynamic fracture may evolve in the future.

\section{The equation of motion for simple cracks}
\label{sec:eom}

What are ``simple'' cracks? Let us define a simple crack as a single crack that propagates along a single (straight) fracture plane~\cite{BFM10}. A simple crack is, basically, a crack that does not undergo any path instabilities. How is the motion of such cracks described? The equation of motion for simple cracks is based on the principle of energy balance \cite{Freund.90,Broberg.1960,Kostrov.1975,Rice.1968}; the energy release rate $G$ (the elastic energy released per unit crack surface) is balanced by the fracture energy $\Gamma$ (the energy dissipation per unit crack surface).

As long as the surrounding material is elastic, $\Gamma$ represents all of the energy dissipated by a crack. For brittle materials, the ``small scale yielding'' assumption is valid; all of the dissipative processes are encompassed in a sufficiently small region surrounding the crack tip \cite{Freund.90}. $\Gamma(v)$ is, in general, a material function that may depend on the crack velocity, $v$ \cite{Irwin.1957,Bergkvist.74,Dally.1979,Kobayashi.1978,Ravi-Chandar.84c,rosakis_freund}. Its $v$-dependence is often unknown, but in some cases can be measured~\cite{Sharon.96,Goldman.10,Baumberger.06c}.

The other ``leg" of the energy balance equation is the energy release rate, $G$. The latter is a property of the elastodynamic solution of a moving crack and can be obtained by the J-integral for steady-state cracks or more generally by the dynamic stress intensity factor. Within the framework of Linear Elastic Fracture Mechanics (LEFM), the near tip stress $\B \sigma$ (as well as displacement gradient) fields are characterized by a universal singularity, $\B \sigma\!\sim\!K_{I,II}/\sqrt{r}$, where the intensity of the singularity is quantified by the stress intensity factors $K_{I,II}$. $K_I$ corresponds to mode I fracture, i.e. tensile cracks, and $K_{II}$ corresponds to mode II fracture, i.e. shear cracks. The stress intensity factors depend on the loading configuration and geometry of a given fracture problem. Once they are known, the energy release rate $G$ can be readily calculated. The equation of motion for a straight crack is then given by $\Gamma(v)\!=\!G$. Next, we briefly discuss two such equations, which were quantitatively tested in recent experiments.

The most well-known equation of motion was obtained for a finite crack propagating in an infinite medium. In this case, the equation takes the form~\cite{Freund.90}
\begin{equation}
  \Gamma (v) = G(l,v)\simeq \frac{\sigma_\infty^2}{E}\, l \left(1-\frac{v}{c_R}\right) \ ,
\label{Freund}
\end{equation}
where a pre-factor of order unity was omitted. Here $l$ is the crack length ($\dot{l}\!=\!v$), $\sigma_\infty$ is the externally imposed tensile stress far from the crack, $E$ is Young's modulus and $c_R$ is the Rayleigh wave-speed. Once $\Gamma(v)$ is known, Eq.~\eqref{Freund} is a first-order nonlinear evolution equation for $l(t)$. It is valid as long as ``small scale yielding'' conditions prevail (i.e. linear elasticity is valid everywhere except for a small region near the tip, the so-called process zone). Additionally, the externally applied stresses should be able to be mapped  to tractions applied to the crack faces and the crack cannot interact with the external boundaries (i.e. it propagates in an effectively unbounded medium).

\begin{figure}
\includegraphics[width=.55\textwidth]{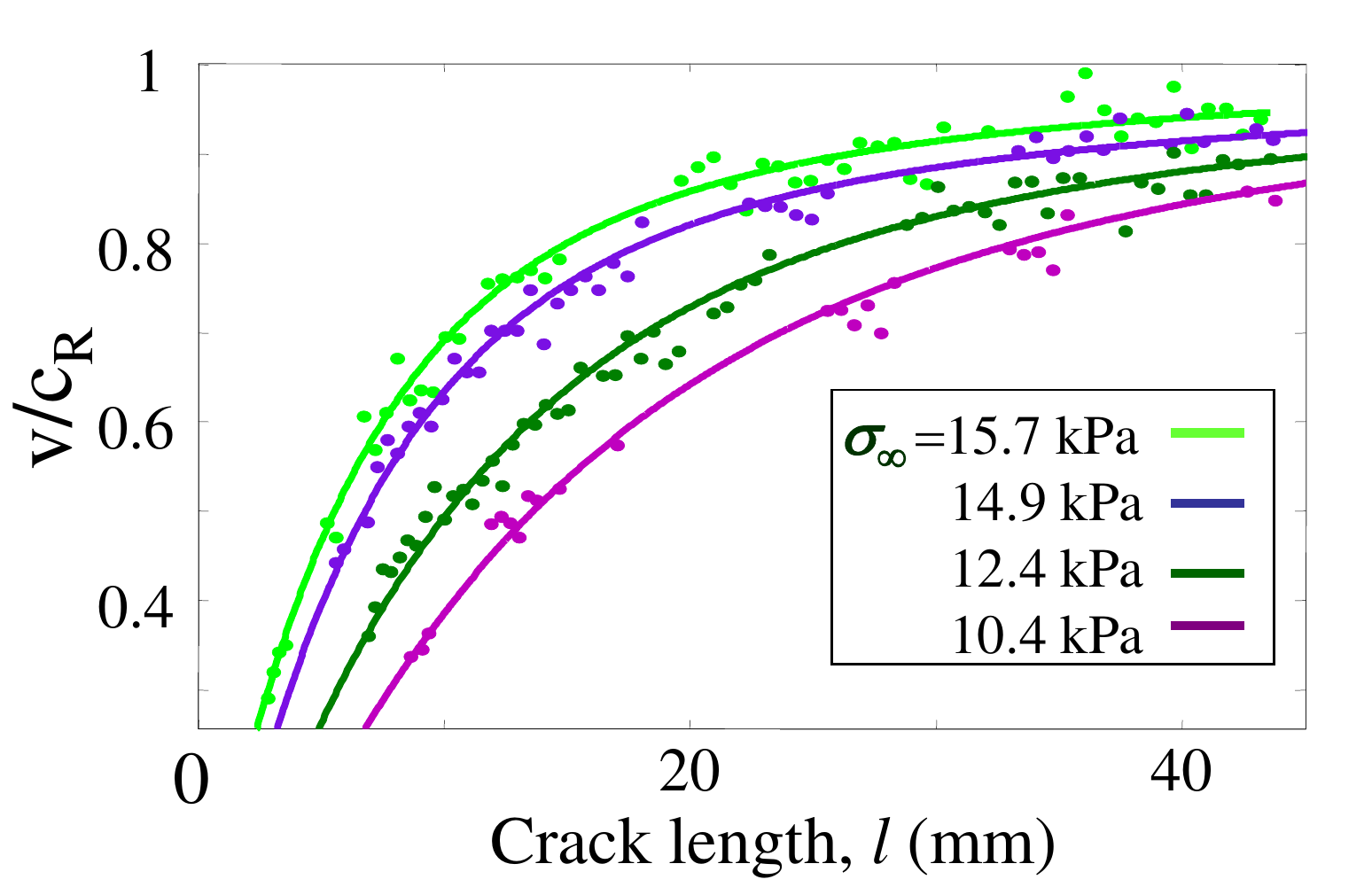}
\caption{The measured crack propagation velocity $v$ as a function of the crack
length $l$ (circles) compared to the LEFM predictions in Eq.~\eqref{Freund} (solid lines),
for various values of the far field tensile stress $\sigma_\infty\!=\!15.7,
14.9, 12.4, 10.4$kPa (top to bottom). All measurements are for single
cracks propagating in effectively ``infinite medium'' conditions ($l(t)\!<\!2b\!=\!125$mm). Adapted from \cite{Bouchbinder.14}.}
\label{equmotion}
\end{figure}

The most interesting properties of Eq.~\eqref{Freund} are that the crack tip dynamics are not inertial, i.e. the tip behaves as a massless ``particle'' where the acceleration $\dot{v}$ plays no role (of course {\em material} inertia is essential for this equation to hold), and that the limiting propagation velocity is $c_R$. The equation has been shown to beautifully
describe crack's motion when its underlying assumptions are met \cite{Bergkvist.74,Washabaugh.94,Sharon.99,Kessler.03,Miller.99}. In particular, this description was shown to be valid in experiments in the brittle acrylic, PMMA, for $v\!<\!0.4c_R$ (when cracks are indeed simple)~\cite{Sharon.99}. More recent experiments showed the equation of motion to be in excellent quantitative agreement with measurements of dynamic cracks in brittle gels for an unprecedented range of velocities, $0\!<\!v\!<\!0.96c_R$, approaching the limiting velocity~\cite{Goldman.10}. This wide velocity range was only accomplished by suppressing crack instabilities. This excellent agreement with theory, presented in Fig.~\ref{equmotion}, was obtained entirely from first principles with no adjustable parameters.

Less known is an equation of motion for simple cracks which is valid when the crack ``knows'' about the external boundaries. While energy balance is always respected, Eq.~\eqref{Freund}  was derived for an system of infinite extent and, therefore, does not account for energy reflected back to the crack from external boundaries. When the waves that the crack emits during propagation are reflected back from the outer boundaries, the crack's dynamics may undergo a qualitative change. Let us consider the important particular case of a crack propagating in a long strip of width $2b$ which is loaded uniaxially by fixed tensile displacements. When the crack is sufficiently long, $l\!\gg\!b$, there is a fixed amount of strain energy per unit area $W$ stored way ahead of the tip and no stored energy well behind the tip. Consequently, under steady state conditions the crack velocity is simply determined by $\Gamma(v)\!=\!W$. Marder derived the crack evolution approaching this steady state to leading order in the dimensionless acceleration $b\dot{v}/c_d^2$, where $c_d$ is the dilatational wave-speed \cite{Marder.91}. In this case, the crack length plays no role (the relevant length scale is $b$, not $l$), but the acceleration (which quantifies the deviation from steady state) is the relevant dynamical quantity. It is worth noting that this acceleration dependence is entirely due to the crack's interaction with the system boundaries and not to, for example, large acceleration corrections to the stress fields  predicted for rapidly accelerating cracks in an infinite system \cite{Liu.1993}.  This (little known) equation of motion takes the form \cite{Marder.91}
\begin{equation}
\label{marder}
\Gamma(v)= G(v, \dot{v}) \!\simeq
W\left[1-\frac{b\,\dot{v}}{c_d^2}\left(1-\frac{v^2}{c_R^2}\right)^{\!-2}\right] \ .
\end{equation}

\begin{figure}
\includegraphics[width=.55\textwidth]{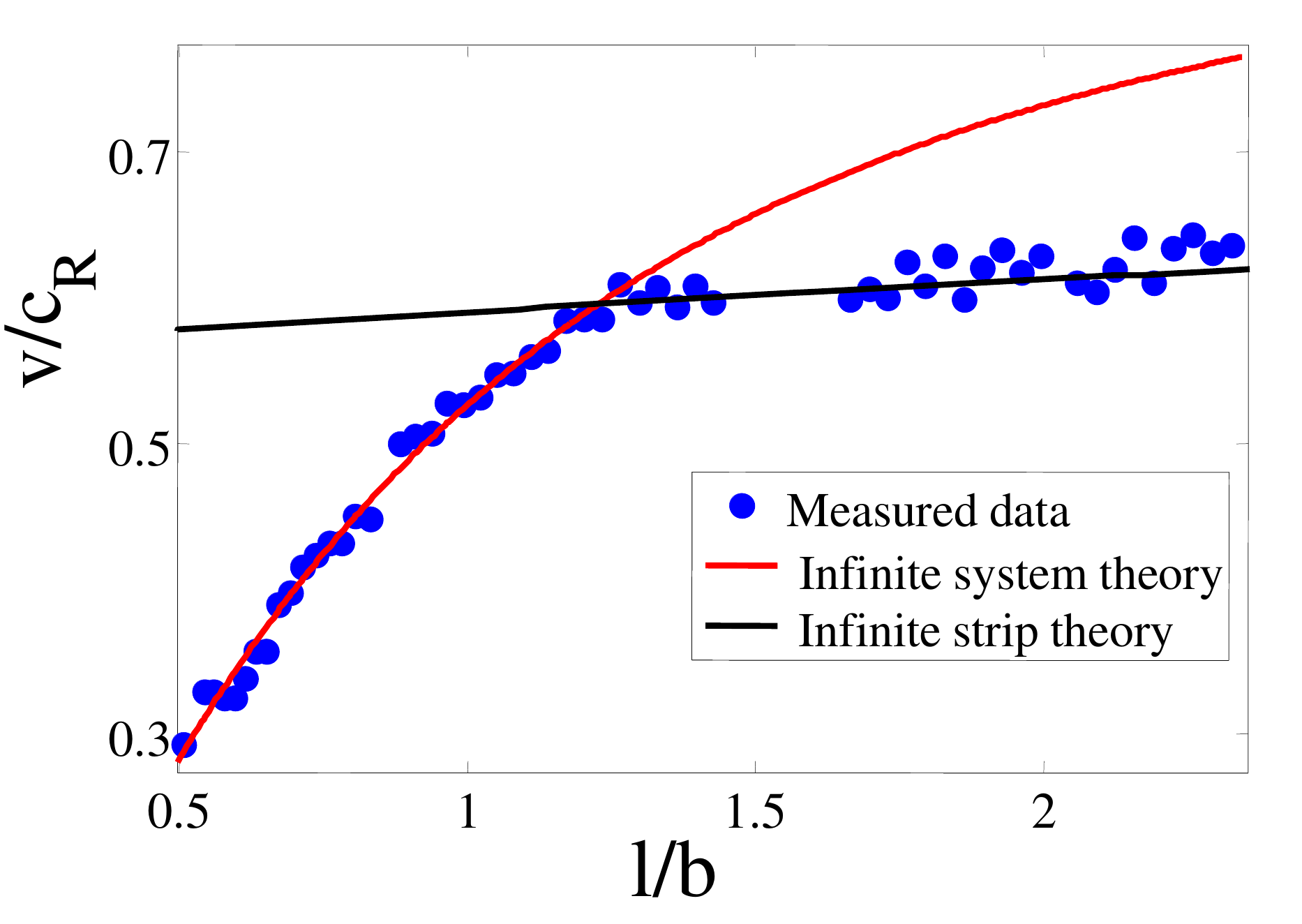}
\caption{The measured crack velocity vs. crack length in a sample of dimensions $120\!\times\!30\!\times\!0.2$mm (circles) compared to the LEFM prediction for an infinite medium (small $l(t)$), Eq.~\eqref{Freund} (red line), and to the LEFM prediction for an infinite strip (large $l(t)$), Eq.~\eqref{marder} (black line). The rather sharp transition between the two regimes occurs at $t\!\sim\!2b/c_s$ ($b$ is the half-width of the strip and $c_s$ is the shear wave-speed), when reflected waves from the vertical boundaries interact with the crack tip. Adapted from \cite{Goldman.10}.}
\label{stripcompare}
\end{figure}

This equation is qualitatively different from Eq.~\eqref{Freund} in one major aspect: the crack tip dynamics are effectively inertial, involving acceleration and an ``effective mass'' proportional to $(1-v^2/c_R^2)^{\!-2}$, rather than being ``massless'' as in Eq.~\eqref{Freund}. These effective inertial dynamics are inherited from the interaction of the crack with the strip boundaries (and hence with its own history, giving rise to ``memory''). Since $W$ is externally controlled and $\Gamma(v)$ is a bounded material function (attaining a finite value as $v\!\to\!c_R$), one can impose $W\!>\!\Gamma(c_R)$. In this case, the divergence of the effective mass proportional to $(1-v^2/c_R^2)^{\!-2}$ as $v\!\to\!c_R$ ensures that $c_R$ is still the limiting velocity as in Eq.~\eqref{Freund} (here $\dot{v}\!\to\!0$ such that $\dot{v}(1-v^2/c_R^2)^{\!-2}$ remains finite as $v\!\to\!c_R$). The predictions of Eqs.~\eqref{Freund}-\eqref{marder} can be tested in a single experiment involving a long strip under fixed displacement boundary conditions. For short times and $l(t)\!\ll\!b$, we expect Eq.~\eqref{Freund} to be valid, while for longer times and $l(t)\!\gg\!b$, we expect Eq.~\eqref{marder} to be valid, with some crossover in between. A comparison to such an experiment, shown in Fig.~\ref{stripcompare}, demonstrates that both Eqs.~\eqref{Freund}-\eqref{marder} are quantitatively correct when valid and that the transition between them is rather sharp.

Therefore, while for many the equation of motion described by Eq.~\eqref{Freund} is unjustifiably thought to be universal, Eq.~\eqref{marder} explicitly demonstrates that this is not the case. In fact, we fully expect that the qualitative dynamics which are inherent in Eq.~\eqref{marder} are in many ways general. If, for whatever reason, a crack is allowed to interact with its past history, we should expect strong memory, inertial-like, effects. Examples where such interactions may occur are when a crack tip/front encounters a localized inhomogeneity or obstacle in its path. In such cases, when the translation invariance along the singular crack front  --- the crack's leading {\em edge} in 3D --- is broken, it has been shown that waves will propagate along the crack's front \cite{Ramanathan.97,Morrissey1998,Morrissey2000,Sharon.01,Sharon.02,Sagy.01,Sagy.04}.  These waves, which transfer energy between different points along the crack front, may influence crack dynamics in much the way that the reflected waves that lead to Eq.~\eqref{marder} impart a crack with inertia. Moreover, experiments have shown that the crack front indeed retains a memory of an obstacle after it has passed it; damped oscillations of the front are observed immediately in the wake of a localized obstacle~\cite{Sharon.02}. These issues will be briefly discussed in Sect.~\ref{sec:3D}.

\section{Weakly nonlinear theory of fracture, near-tip fields and intrinsic length scales}
\label{sec:wnfm}

The equations of motion for simple cracks, which were discussed above, are based on energy balance. We would like now to consider the actual fields that mediate the transport of energy to the crack tip region and may control the fracture process itself. LEFM predicts that sufficiently close to the crack tip the stress field features a universal square-root singularity, $\B \sigma\!\sim\!1/\sqrt{r}$, where the pre-factor incorporates a universal tensorial function of the angle $\theta$ and the propagation velocity $v$ ($r$ and $\theta$ are polar coordinates co-moving with the tip), and the scalar multipliers --- the stress intensity factors. Due to linearity, the displacement gradients feature the same singularity $\nabla \B u^{(1)}\!\sim\!1/\sqrt{r}$, where $\B u^{(1)}$ is the linear elastic displacement field.

The LEFM fields possess certain peculiar features that are worthwhile noting. For example, in mode I (tensile) fracture, the crack parallel component of the stress tensor ahead of the tip, $\sigma_{xx}(r,\theta\!=\!0,v)$, is {\em larger} than the tensile component $\sigma_{yy}(r,\theta\!=\!0,v)$ for every finite crack velocity $v\!>\!0$~ \cite{Rice.68,Langer.98,Fineberg.99}. In fact, the ratio $\sigma_{xx}(r,\theta\!=\!0,v)/\sigma_{yy}(r,\theta\!=\!0,v)$ diverges as $v\!\to\!c_R$. This feature implies that it is not easy to explain why mode I cracks propagate perpendicularly to the tensile loading ($y$) axis based on the LEFM universal fields. While this feature has been recognized in the past, it has not received much attention. A corollary of this is that there always exists a crack propagation velocity $v_0\!=\!\tfrac{1}{2}(\sqrt{c_d^2+8c_s^2}-c_d)$ (cf. Eq. (49) in~\cite{Bouchbinder.14} and note that $c_s$ is the shear wave-speed) above which $\pa_y u^{(1)}_y(r,\theta\!=\!0,v)$ becomes {\em negative}. That is, the universal LEFM fields predict that tensile cracks, which ultimately involve extending --- and eventually breaking --- bonds ahead of their tips, propagate with a predominantly compressive displacement gradient ahead of their tip \cite{Livne.10}.

These features suggest that the LEFM universal fields are insufficient to fully explain the fracture process, which is supported by explicit measurements of the strain fields surrounding dynamic cracks~\cite{Livne.08}. This by no means implies that LEFM is wrong, this theory accurately predicts the flux of energy into the tip region and the universal LEFM fields pose the proper boundary conditions on any inner problem which is defined inside the process zone. It does imply, though, that the {\em length scale} at which these fields are valid is quite distinct from the scale where fracture actually takes place. LEFM, however, is a scale-free theory that provides no specific information about where it is valid and on what length scale it breaks down. This issue of the absence of an intrinsic length scale may have serious consequences in the context of dynamic instabilities, as will be discussed below.

Recently, these issues have been addressed in the framework of the ``Weakly nonlinear theory of fracture" \cite{Bouchbinder.14,Bouchbinder.08,Bouchbinder.09,Bouchbinder.2010,Bouchbinder.2010_IJF}. This theory is based on a general expansion of the strain energy functional $U$ in the form
\begin{equation}
\label{U_E}
U(\B E)\!\simeq\!\frac{\lambda}{2} \left(tr\!\B E\right)^2 + \mu ~tr\!\B E^2 + \beta_1 \left(tr\!\B E\right)^3 + \beta_2 tr\!\B E~ tr\!\B E^2 + \beta_3 tr\!\B E^3 + \C O(\B E^4) \ ,
\end{equation}
where $\B E \!=\! \tfrac{1}{2}[(\nabla \B u) + (\nabla \B u)^T + (\nabla \B u)^T (\nabla \B u)]$ is the nonlinear Green-Lagrange strain tensor \cite{Holzapfel.00} and $\B u$ is the displacement field. Here $\lambda$ and $\mu$ are the standard Lam\'e (second order) constants and $\{\beta_1, \beta_2, \beta_3\}$ are the third order elastic constants~\cite{51Murnaghan}, which are basic physical quantities that are related to the leading anharmonic contributions to the interatomic interaction potential. The third order elastic constants are not free parameters, but rather quantities that are either calculated from a fully nonlinear elastic energy functional, if known, or measured directly in experiments \cite{Bouchbinder.14}.

As is evident from Eq.~\eqref{U_E}, the weakly nonlinear theory takes into account the leading elastic nonlinearities when LEFM breaks down. The basic idea is that in the presence of the intense displacement gradients near crack tips in brittle materials, linear elasticity first gives way to nonlinear elasticity, rather than to dissipation. As such, it is an asymptotic (in space, where LEFM breaks down near the tip of a crack) and perturbative (in the magnitude of displacement gradients) theory based on the expansion $\nabla \B u\!=\!\nabla \B u^{(1)} + \nabla \B u^{(2)} + {\C O}(|\nabla \B u^{(3)}|)$, where the superscripts denote different orders in the displacement gradients. $\B u^{(1)}$ simply corresponds to LEFM and $\B u^{(2)}$ satisfies a new equation that depends both on the nonlinearity and $\B u^{(1)}$. The solution to this equation takes the form (see details in \cite{Bouchbinder.14})
\begin{equation}
u_x^{(2)} \sim \frac{K_{I,II}^2}{\mu^2}\left[\log{r} + f_x^{I,II}(\theta,v)\right], \quad u_y^{(2)} \sim \frac{K_{I,II}^2}{\mu^2} f_y^{I,II}(\theta,v)
\label{WNFM}
\end{equation}
and is inherently mode I (tensile) for $\B u^{(1)}$ corresponding to {\em both} mode I {\em and} mode II. That is, the explicitly calculable functions $f_{x,y}^{I,II}(\theta,v)$ satisfy $f_x^{I,II}(\theta,v)\!=\!f_x^{I,II}(-\theta,v)$ and $f_y^{I,II}(\theta,v)\!=\!-f_y^{I,II}(-\theta,v)$. Note that $r$ inside $\log{r}$ in Eq.~\eqref{WNFM} should be made non-dimensional, but this simply adds a constant and is not included here.

\begin{figure}[here]
\includegraphics[width=0.85\textwidth]{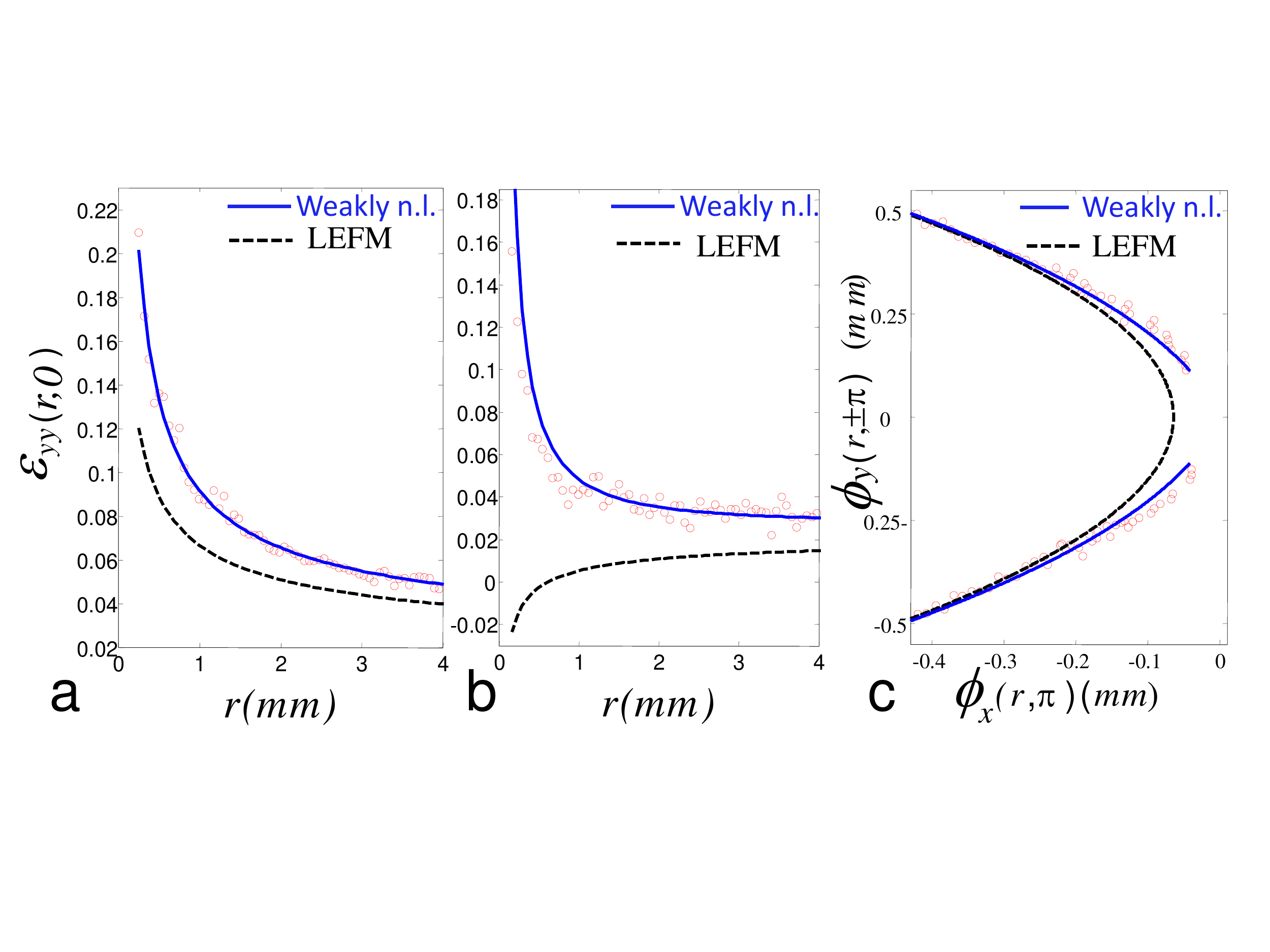}
\caption{(a) The measured $yy$-component of the displacement gradient tensor $\varepsilon_{yy}(r,\theta\!=\!0)\!=\!\pa_y u_y(r,\theta\!=\!0)$ ahead of a crack propagating at $v\!=\!0.20c_s$ (circles) compared to the LEFM prediction (dashed line) and the weakly nonlinear theory prediction (LEFM supplemented with Eq.~\eqref{WNFM}) (b) The same for $v\!=\!0.78c_s$. Note, in particular, that for $v\!>\!v_0=0.73c_s$, LEFM predicts $\varepsilon_{yy}(r,\theta\!=\!0)\!<\!0$ sufficiently close to the tip, while both the experiment and the weakly nonlinear theory show that $\varepsilon_{yy}(r,\theta\!=\!0)\!>\!0$, as is physically expected in mode I fracture. (c) The measured crack tip profile ($\phi_y(r,\theta\!=\!\pm\pi)$ vs. $\phi_x(r,\theta\!=\!\pi)$ for $v\!=\!0.53c_s$, where $\B \phi(\B x)\!\equiv\!\B x+\B u(\B x)$) (circles) compared to the best fit of
the parabolic LEFM profile (dashed line) and that
predicted by the weakly nonlinear theory (solid line). The effect of the logarithmic stretching term in Eq.~\eqref{WNFM} is evident as the tip is approached. Adapted from~\cite{Bouchbinder.08}, where additional details can be found.
}
\label{fig:WNFM}
\end{figure}

The theory, as is clearly observed in Eq.~\eqref{WNFM}, predicts a stronger than $1/\sqrt{r}$ divergence of the displacement gradient tensor field, $\nabla \B u^{(2)}\!\sim\!1/r$, and a logarithmic stretching of the crack tip opening displacement, which is strictly parabolic in LEFM~\cite{Freund.90}. This $1/r$ displacement gradient singularity has a fundamentally different status compared to apparently similar singularities in LEFM \cite{Hui.95,Williams.57}. In particular, its contribution to the J-integral vanishes identically. This, and other properties, are discussed in detail in ~\cite{Bouchbinder.14,Bouchbinder.09}. These predictions were quantitatively verified by direct experiments on rapid cracks, as shown in Fig.~\ref{fig:WNFM}. In particular, it is explicitly shown (panel b) that for a propagation velocity for which LEFM predicts a negative displacement gradient $\pa_y u_y^{(1)}(r,\theta\!=\!0, v)\!<\!0$ near the tip (i.e. $v\!>\!v_0$, where in this case $v_0\!=\!0.73c_s$), both the experimentally observed and theoretically predicted $\pa_y u_y(r,\theta\!=\!0, v)$ are positive, the latter {\em directly} due to the nonlinear contribution $\pa_y u_y^{(2)}(r,\theta\!=\!0, v)$.

To sum up, the weakly nonlinear theory of fracture --- and its experimental verification --- show that elastic nonlinearities give rise to a stronger near tip singularity compared to LEFM and to an extensional displacement gradient ahead of the tip, $\pa_y u_y(r,\theta\!=\!0,v)\!>\!0$, even at propagation velocities where LEFM predicts otherwise (i.e. for $v\!>\!v_0$, cf. Fig. \ref{fig:WNFM}b). The latter at least recovers the basic physical intuition that mode I fracture occurs under extensional deformation ahead of the propagating tip. As was mentioned above, when nonlinearities are taken into account for mode II cracks \cite{Harpaz.12}, a symmetry-breaking effect in which LEFM mode II (shear) cracks feature mode I (tensile) properties naturally emerges in this theoretical framework~\cite{Stephenson.82}.

As the weakly nonlinear theory features two different singularities associated with two different constitutive relations, it predicts the existence of a new intrinsic (i.e. non-geometric) length scale, $\ell_{nl}$, which characterizes the scale at which the two singular solutions are comparable in magnitude, i.e. roughly when $|\nabla \B u^{(1)}|\!\simeq\!|\nabla \B u^{(2)}|$. This intrinsic length scale is a dynamic quantity (i.e. it exhibits a non-trivial velocity dependence) that depends on the third order elastic constants. Its length dimension is inherited from $\Gamma/E$, i.e. $\ell_{nl}\!\propto\!\Gamma/E$, with a nontrivial and not necessarily order unity pre-factor. Finally, we note that very recently an Extended Finite Element Method (XFEM) analysis of nonlinear fracture problems based on the weakly nonlinear theory has been performed~\cite{Rashetnia.15}. It was shown that using the analytic properties of the solution in Eq.~\eqref{WNFM} to construct crack tip enrichment functions provides an accurate and efficient tool for solving finite strain fracture problems.

The weakly nonlinear theory of fracture, which invokes three directly measurable third order elastic constants, provides a general and systematic way to go beyond LEFM at smaller scales near the tip. It naturally resolves apparently paradoxical physical features of the LEFM universal fields and gives rise to a new intrinsic length scale, which is missing in LEFM. As will be discussed soon, the latter plays an important role in dynamic instabilities.

\section{Crack instabilities}
\label{sec:Instabilities}

\subsection{The micro-branching instability}
\label{microbranches}

Historically, early measurements showed that a running crack did not seem to be described by the equations of motion described in Sect.~\ref{sec:eom}. In particular, these measurements suggested that even as robust a prediction as the limiting velocity of a dynamic crack seemed to be off; cracks in brittle materials rarely surpassed even 60\% of $c_R$. These observations  suggested that something was seriously wrong with our fundamental understanding of rapid crack dynamics, but it was unclear where the root of the problem lay. One could always ``blame'' the velocity dependence of the fracture energy, $\Gamma(v)$, but then one had to explain why, in brittle amorphous materials $\Gamma(v)$ needed to increase precipitously in order to prevent Eq.~\eqref{Freund} from allowing a crack to approach $c_R$. Even more puzzling was the fact that this behavior occurs in a variety of materials whose molecular dissipative mechanisms are expected to be wholly unrelated; these materials ranged from brittle (cross-linked and uncross-linked) polymers to glasses.

The answer to the apparent paradox suggested above is that at high velocities, the crack is no longer a ``simple crack''. Instead, as first observed in \cite{Ravi-Chandar.84c,Ravi-Chandar.84b}, a crack's structure at relatively high propagation velocities is entirely different from that envisioned in the picture of simple crack;  high-speed photographs of rapid cracks showed that the structure of the crack tip was anything but simple. Instead of the propagation of a single simple crack, these experiments showed that rapid cracks are composed of an ensemble of simultaneously propagating cracks.  Later experiments~\cite{Fineberg.91}, suggested that the break-up of a simple crack to multiple cracks occurred at a critical velocity, at which a single crack became unstable. This instability, which became known at the ``micro-branching'' instability, was later shown to be characteristic of many brittle amorphous materials~\cite{Fineberg.99,Ravi-Chandar.84c,Fineberg.91,Cramer.00,Sharon.98,Sharon.2.96,Gross.93,Sharon.95,Boudet.95,Livne.05,Yang.96,Ravi-Chandar.97}. A comprehensive review of some of the early work on the micro-branching instability can be found in~\cite{Fineberg.99}.
\begin{figure}
\includegraphics[width=.7\textwidth]{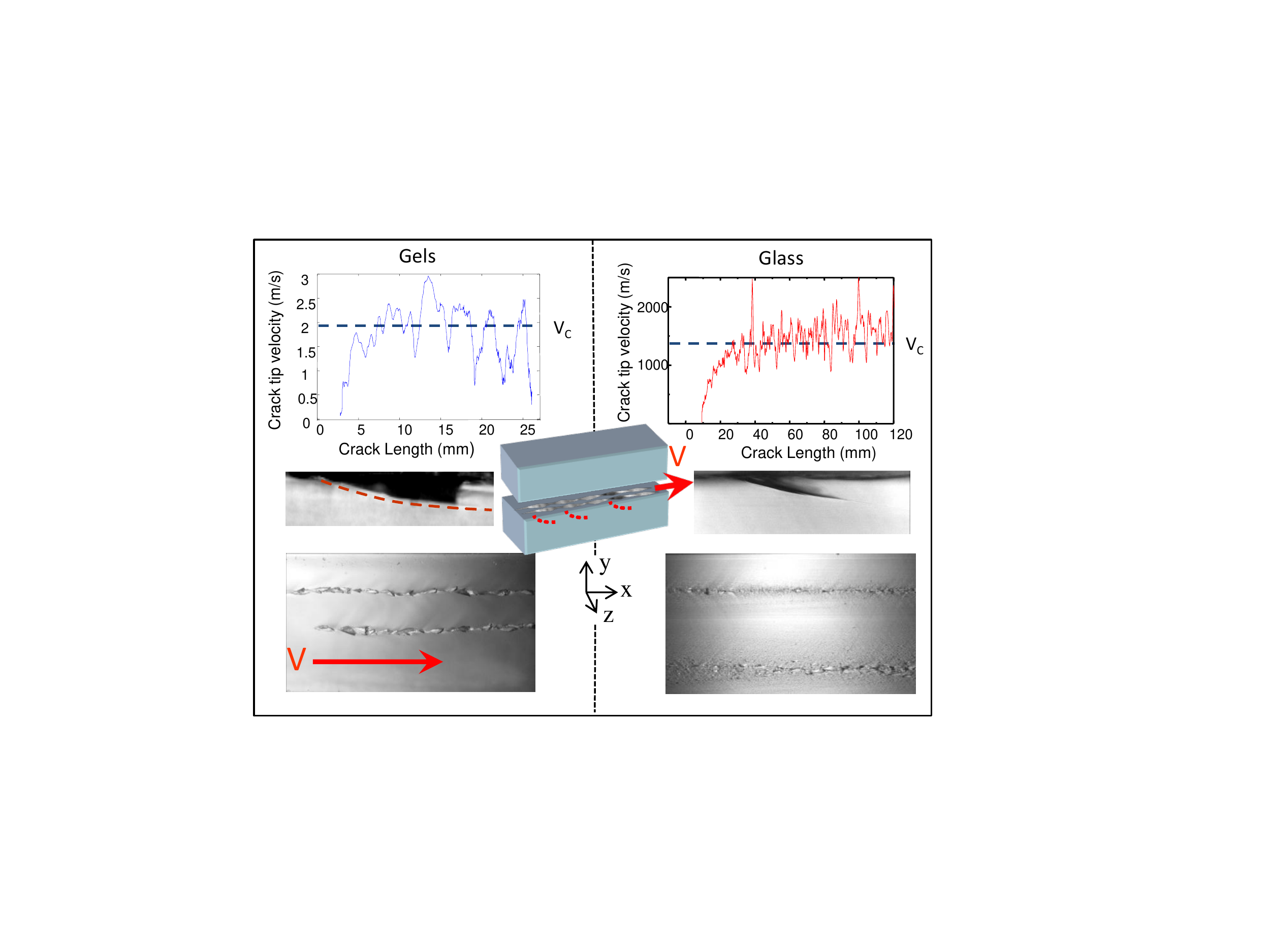}
\caption{ A comparison between the instability in both soft polyacrylamide gels (left) and soda-lime glass (right). Once a crack attains a critical velocity of $v\!\simeq\! 0.4c_R$, a single ``simple'' crack may become unstable to the micro-branching instability. At this point (top) the instantaneous velocity of the crack undergoes violent oscillations that correspond to the formation of subsurface frustrated micro-branches (center), where side views ($xy$ plane) of micro-cracks in both materials are presented ($x$ is the propagation direction, $y$ is the loading direction and $z$ is the direction along the crack front). Micro-branches are very similar in both materials, characterized by a power-law functional form. (bottom) Photographs of the resulting fracture surface ($xz$ plane) formed by cracks propagating from left to right at $v\!\simeq\!0.5c_R$. The chains of structures on the fracture surface correspond to chains of micro-branches that are aligned in the propagation direction $x$ and highly localized in the $z$ direction. The vertical dimensions of the photographs are about $0.5mm$. The cartoon in the center describes the geometry; the $xz$ fracture surface is formed by a propagating cracks whereas  micro-branches (dotted lines) extend below the surfaceAdapted from \cite{Bouchbinder.14}.}
\label{instabilities1}
\end{figure}

\begin{figure}
\includegraphics[width=.55\textwidth]{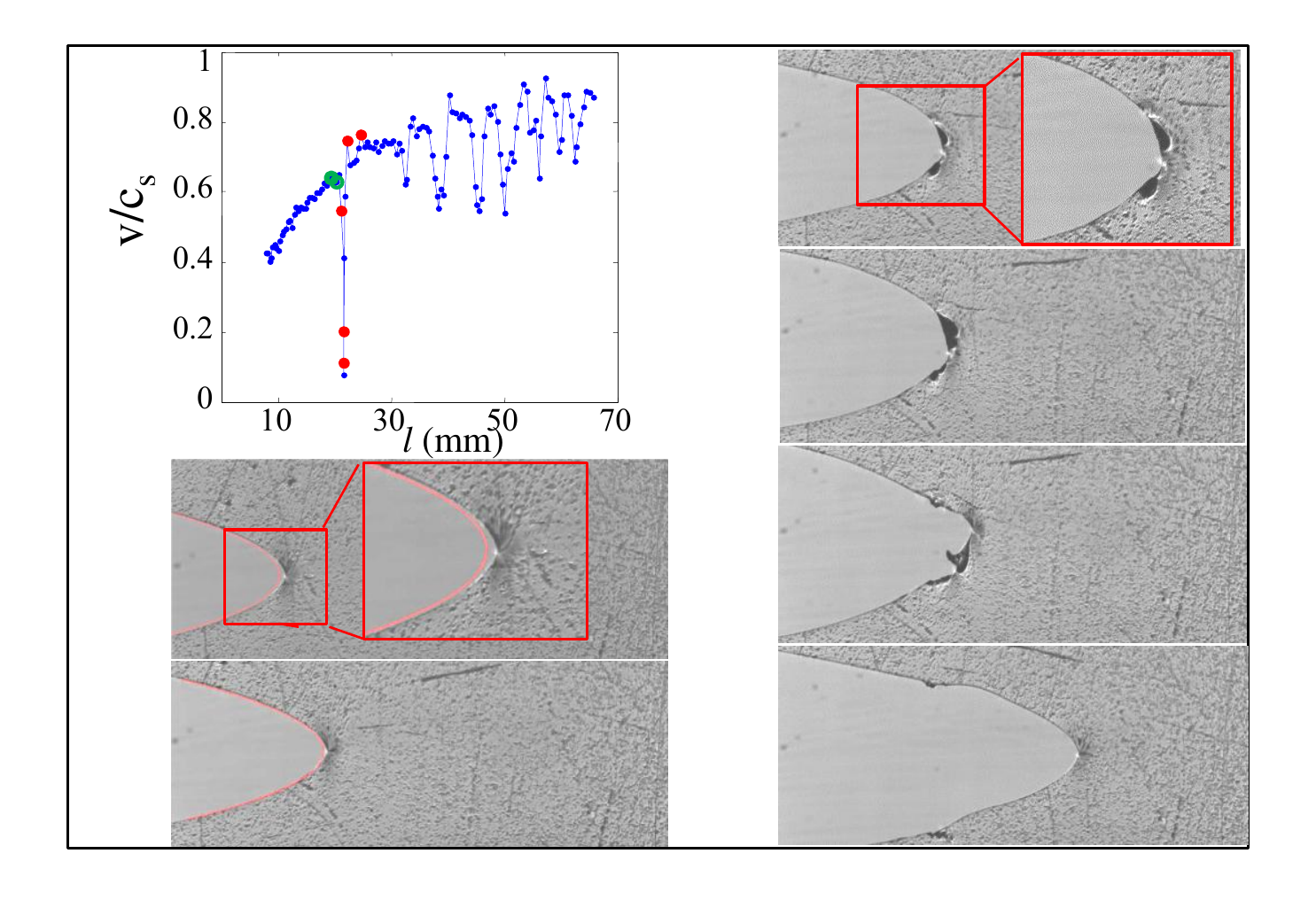}
\caption{ A series of photographs of the quasi-parabolic profile of running cracks in polyacrylamide gels. The two lower left panels depict the parabolic crack tip opening profile of ``simple" cracks propagating just prior to the onset of the micro-branching instability (corresponding to the green dot on the velocity vs. crack length figure in the upper right corner). Right panels: 4 profiles of cracks undergoing the micro-branching instability, corresponding to the red dots in the velocity measurements. Note the small micro-cracks formed at the tip of the main crack. Adapted from \cite{Bouchbinder.14}.}
\label{instabilities2}
\end{figure}

What is the micro-branching instability? For low crack velocities, the crack propagates smoothly and the fracture surfaces formed are smooth and mirror-like, with the only apparent structure caused by material defects interacting with the singular front \cite{Yang.96,Scheibert.10}. At a critical velocity of $v_{mb}\!\simeq\!0.4c_R$, the crack undergoes a dynamic instability. As shown in Fig. \ref{instabilities1}, rapid fluctuations appear in the instantaneous crack velocity and, at the same time, non-trivial structure emerges on the fracture surface. The fracture surface structure is accompanied by frustrated crack-branching, as multiple microscopic crack branches (``micro-branches'') are observed below the fracture surface.  As the crack velocity increases beyond $v_{mb}$, both the length and density of the micro-branches increase with $v$. The increased length and density of the micro-branches gives rise to correspondingly rougher fracture surface features, larger velocity fluctuations, and a sharp increase of the fracture energy that is proportional to the total micro-branch length \cite{Sharon.2.96,Sharon1996}. As Fig.~\ref{instabilities1} demonstrates, these characteristic features are independent of the brittle material; shown are two extremely different classes of material, poly-acrylamide gel (an aqueous elastomer) and soda-lime glass which exhibit identical characteristic behavior~\cite{Livne.05}. Note the use of brittle gels has enabled experiments to probe the fracture process in unprecedented detail, as demonstrated in Fig.~\ref{instabilities2}.

Over the past two decades, there have been a number of notable attempts to describe the micro-branching instability by supplementing or extending LEFM in various ways. These include phase-field models \cite{Aranson.00,Karma2004,Henry2004,Henry.08,Spatschek.06,SBK11,Henry.13}, phase transformation models \cite{Brener.11}, cohesive-zone models \cite{Miller.99,Langer.98,Falk.01}, lattice models~\cite{Holland.99,Slepyan.02,Marder.93.prl,Marder.95.jmps,Heizler.02,Kessler.01,Guozden.09}, models based on the ``Principle of Local Symmetry'' \cite{Adda-Bedia.99,Movchan.05,Bouchbinder.07}, energetic bounds on crack branching \cite{Eshelby.70,Adda-Bedia.05,Bouchbinder.05a} and models based on non-linear constitutive behavior near the crack tip \cite{Gao.96,Buehler2006,Bouchbinder.09b}. Although many of these models exhibited some properties that were qualitatively consistent with the experimental observations, they are inherently deficient because they focussed on 2D (i.e. they actually studied macro-branching, which appears as a tip-splitting instability). As will be shown below, micro-branching appears to be an intrinsically 3D instability that cannot be satisfactorily described as a 2D phenomenon.

On the conceptual level, we note that most of previous works have considered the near-tip region as a passive region that both regularizes the singular fields predicted by LEFM and accounts for the dissipative processes at the tip, but otherwise plays no other active role in the dynamics. As will be shown next, recent developments may suggest that understanding crack instabilities may entail the introduction of new physical ingredients (e.g. length scales) in which the near-tip region plays a more active role.

\subsection{The oscillatory instability}
\label{oscillations_section}

By manipulating the geometry and thickness of the brittle gels mentioned above, it is possible to statistically suppress the micro-branching instability in sufficiently thin samples ($200-300\mu$m)~\cite{Livne.07,Goldman.12}; while micro-branching can occur for {\em all} velocities $0.3c_s\!<\!v\!<\!0.9c_s$, the probability of exciting them decreases for both thin samples and large crack accelerations. When micro-branching is suppressed, simple brittle cracks can reach extreme velocities. In such cases single crack dynamics, as described by LEFM (Fig.~\ref{equmotion}), can be obtained for velocities approaching $c_R$.

Can these cracks actually reach their asymptotic velocity? It turns out that this is prevented by a {\em new} instability. Experiments in gels have found that at velocities surpassing about 90\% of the shear wave-speed, $c_s$, a straight crack becomes unstable to sinusoidal path oscillations \cite{Livne.07}. This instability is particularly intriguing since it involves a finite instability wavelength at onset. The value of the observed wavelength was not dependent on either the system geometry or the loading conditions. This suggests the existence of an intrinsic scale that can not exist in linear elastic solutions for cracks, which are scale-free.
\begin{figure}[here]
\includegraphics[width=.7\textwidth]{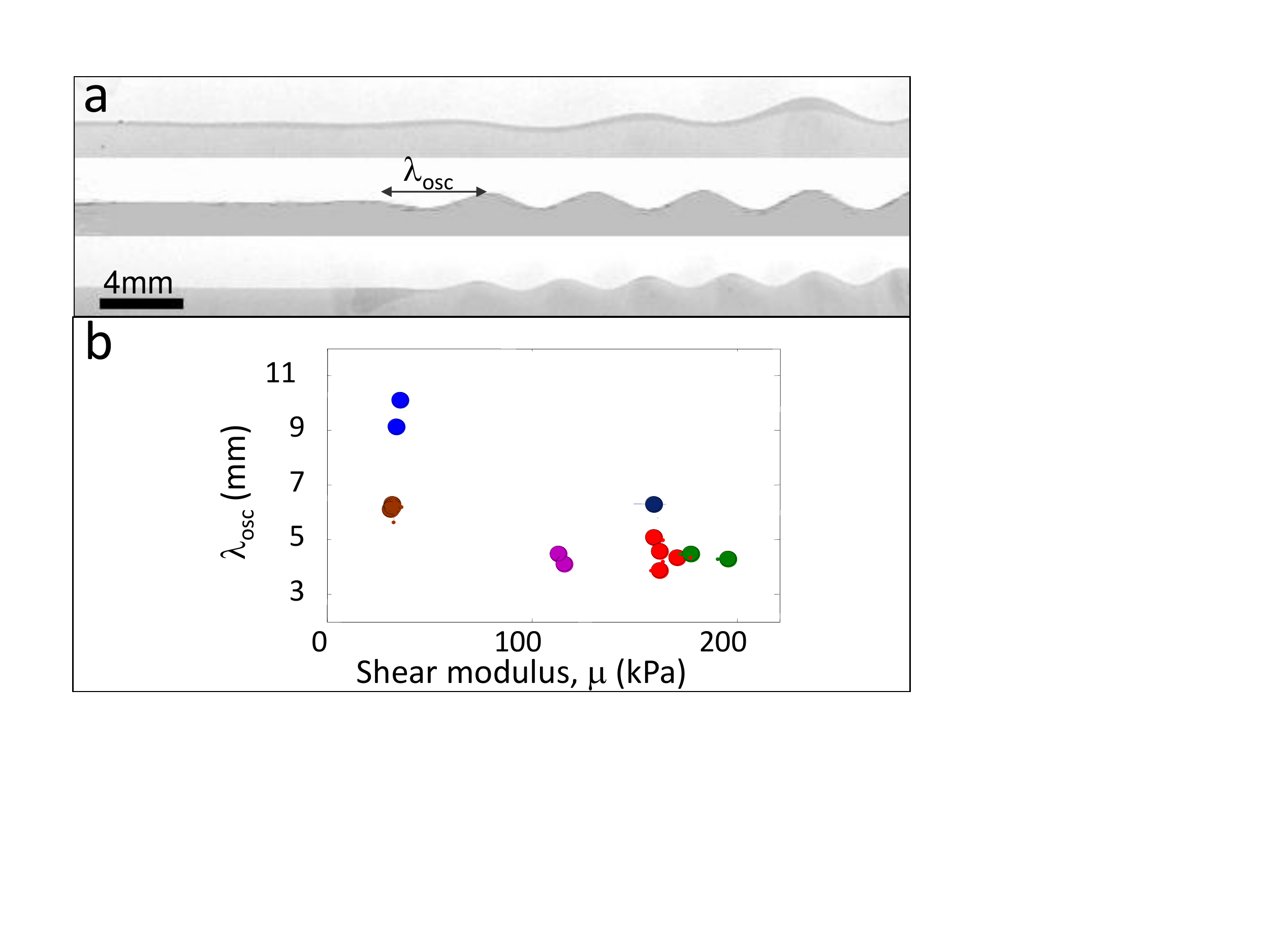}
\caption{(a) Typical photographs of the $xy$ profiles of fracture surfaces at the onset of the oscillatory instability; from top to bottom: gels with shear moduli $\mu\!=\!36$kPa, $143$kPa, $168$kPa.
(b) The oscillation wavelength, $\lambda_{osc}$ changes significantly with the material, as characterized by $\mu$. Symbol colors as in (b). Adapted from \cite{Goldman.12}.}
\label{oscillations}
\end{figure}

This oscillatory instability should be distinguished from apparently similar oscillatory cracks observed in rubber \cite{Deegan.02.rubber}. The oscillatory cracks in rubber were intersonic ($c_s\!<\!v\!<\!c_d$)~\cite{Marder.JMPS.2006},  entailed bi-axial loading, and were subjected to very large background strains (which make the material response nonlinear everywhere). The  oscillatory cracks in the elastomer gels were subsonic and were observed under uniaxial loading of such magnitude ($\sim\!10\%$) that the deformation away from the tip was predominantly linear elastic.

\subsection{Understanding 2D instabilities}
\label{sec:2Dinstabilities}

In Sects.~\ref{sec:eom} and \ref{sec:wnfm} we discussed simple tensile cracks that propagate along straight paths in effectively 2D bodies. By so doing, we excluded spontaneous symmetry-breaking instabilities from the discussion. Such dynamic instabilities, however, are rather prevalent and pose some of most intriguing open questions in the field of dynamic fracture. In sections \ref{microbranches} and \ref{oscillations_section} we described two different instabilities that take place in fast fracture.

The oscillatory instability described in Fig.~\ref{oscillations} is, in a sense, the simpler of the two instabilities. Once excited, the oscillatory instability breaks the translational invariance of a crack in the propagation ($x$) direction. The fracture surface created during the crack's motion is wavy but mirror-like; there is no apparent structure along the fracture surface. Thus, we can consider the crack front to be translationally invariant along the direction transverse ($z$ direction) to the crack's propagation direction. It is, therefore, sufficient to idealize the crack tip as an oscillating point within an effectively 2D medium.

In contrast to the oscillatory instability, when the micro-branching instability is excited, distinct structures are generated along the fracture surface that are highly localized along both the $x$ and $z$ directions (see, for example, the lower panels of Fig.~\ref{instabilities1}). As the crack front is {\em not} translationally invariant along the $z$ direction, it may not be sufficient to consider a crack propagating within a 2D medium to understand micro-branching.

We now consider the oscillatory instability in a 2D medium. As mentioned above, the most interesting physical property of this instability, beyond its existence, is that its wavelength does not scale with the geometric dimensions of the sample. This implies that it cannot be explained by LEFM, as it possesses no intrinsic length scale beyond geometric length scales. What is then the origin of this instability and the emerging intrinsic length scale?

To address this question, and dynamic fracture instabilities in a broader context, we need to consider first a basic aspect of the current theoretical status of the field. When cracks propagate along straight paths (or other predetermined paths), the quantity of interest is the rate of crack growth, which involves a scalar relation based on energy balance. In the most general case, however, the path is an intrinsic property of the dynamic fracture problem and is selected self-consistently as in a large class of moving free-boundary problems (see, for example,~\cite{RBfronts,Saarloos,CrossHohenberg,lightning,Sivashinsky,KPZ}). Consequently, we need an equation of motion for crack tips (in 2D) and crack fronts (in 3D) that will allow the determination of both the direction of propagation and its rate. Such equations are intrinsically non-scalar in nature and involve physics that go significantly beyond energy balance alone. In particular, such equations are essential in the context of instabilities as without them one cannot even mathematically pose the question of stability, which amounts to perturbing solutions of existing and well-defined differential equations. We currently lack well-established crack tip/front equations of motion for general, non-straight, dynamics.

In 2D and under quasi-static conditions, $v\!\ll\!c_R$, it has been shown that the Principle of Local Symmetry  \cite{Goldstein.Salganik} offers good approximations for crack paths \cite{Cotterell.Rice,Adda-Bedia.95,Bouchbinder.03,Pham.08,Corson.09,hakim2005crack,Hakim.09}. This approximation suggests that under quasi-static conditions, and under the assumption that no intrinsic length scales exist near the tip, crack propagation along smooth paths will annihilate any shear (mode II) component of the fields ahead of the tip. Hence, this path selection criterion proposes that conditions of local tensile symmetry are almost maintained at a crack's tip, which mathematically amounts to $K_{II}/K_I\!\ll\!1$. As the stress intensity factors are generically non-local functionals of the path, the latter is in fact an integro-differential equation. At present, to the best of our knowledge, there is no evidence that the Principle of Local Symmetry offers good approximations also under fully dynamic conditions. In fact, some recent experiments might indicate otherwise~\cite{Goldman.15}.

Recently, a dynamic crack tip equation of motion, which takes into account the existence of an intrinsic length scale $\ell_{nl}$ and the possible high propagation velocities, was proposed~\cite{Bouchbinder.09b}. The basic idea is that in the presence of a finite length scale $\ell_{nl}$ near the tip, causality implies a time delay between the singular fields driving the crack (as described by LEFM) and the actual crack tip. Taking into account this time delay along with scaling arguments and symmetry considerations~\cite{Hodgdon.93}, a dynamic crack tip equation of motion has been constructed~\cite{Bouchbinder.09b}.

\begin{figure}
\includegraphics[width=.75\textwidth]{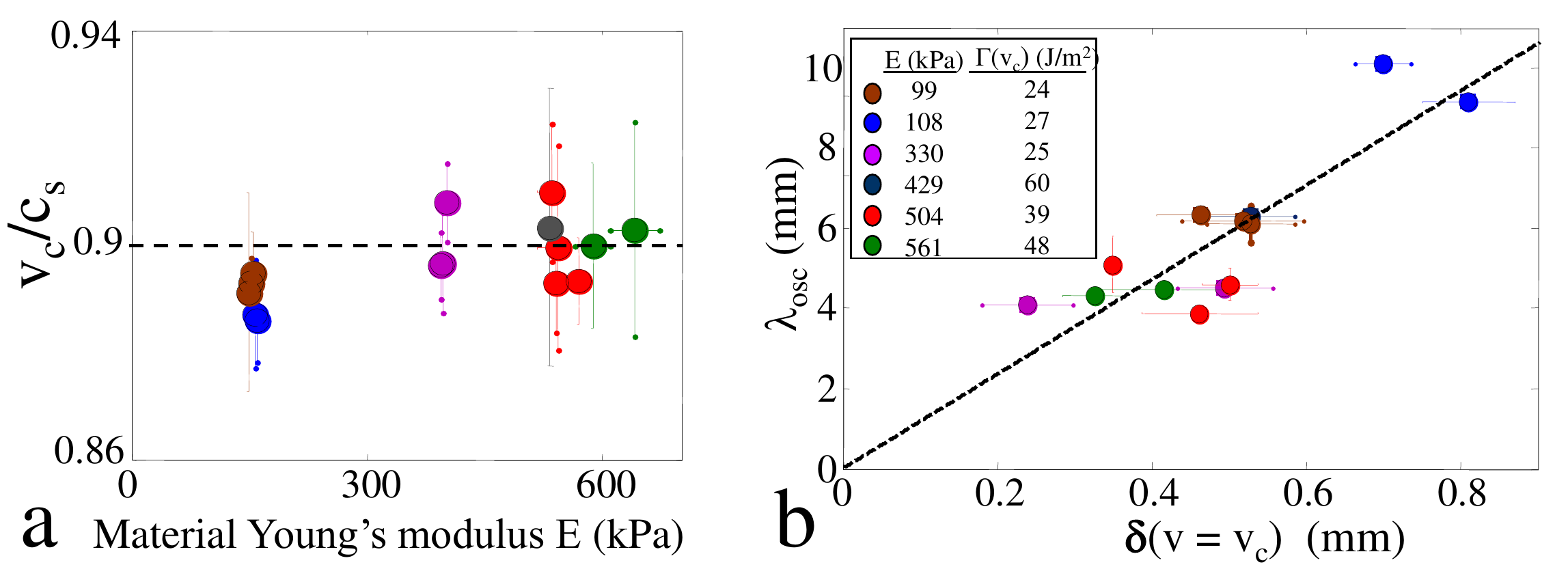}
\caption{(a) The scaled critical velocity for the onset of the oscillatory instability, $v_c/c_s\!\approx\!0.9$, is constant. $v_c$ is defined as the maximal velocity prior to the instability onset in each material. Symbol colors correspond
to the legend in (b). (b) Comparison between an experimental measure of the length scale associated with elastic nonlinearities $\delta(v_c)$ (measured at the critical velocity $v_c$, see~\cite{Goldman.12} for details) and the oscillation wavelength $\lambda_{osc}$. Note that the different combinations of $\mu$ and $\Gamma$ are used to produce these 15 independent measurements. The dashed line is a
guide to the eye. Adapted from~\cite{Goldman.12}.}
\label{oscillations_comparison}
\end{figure}

This equation was used to study the linear stability of straight crack propagation against sinusoidal perturbations, with the help of the Willis-Movchan perturbation theory for the stress intensity factors \cite{Willis_Mochvan.95,Willis.97,Obrezanova.02a}. This analysis predicted an oscillatory instability of mode I cracks to infinitesimal out-of-plane perturbations above a high propagation velocity of $v_c\!\simeq\!0.8c_s$, with the following properties: (i) The normalized critical velocity $v_c/c_s$ is nearly universal (it exhibits a weak dependence on Poisson's ratio $\nu$) (ii) The wavelength of oscillation is proportional to the intrinsic length scale $\ell_{nl}\!\propto\!\Gamma(v_c)/E$ \cite{Goldman.12}. These predictions were recently tested by extensive experiments in which $\Gamma$ and $E$ were varied over a substantial range \cite{Goldman.12}. As Fig. \ref{oscillations_comparison} shows, a favorable agreement between theory and experiments was demonstrated.

We believe that these results are important because they show that the process zone is not a passive region responsible only for nonlinearity and energy dissipation, but rather an active region whose accompanying length and time scales can significantly affect macroscopic fracture dynamics, in particular crack instabilities. The importance of a length scale proportional to $\Gamma/E$ has also been highlighted in experiments on the quasi-static fracture of hydrogels, where it was shown to play a decisive role in a mode I cross-hatching instability \cite{Baumberger.08}, in a wetting-induced branching instability \cite{baumberger.10} and in a mixed-mode I+III (mode III corresponds to anti-plane shear fracture)  \'echelon instability \cite{Ronsin.14}. We suspect that the incorporation of intrinsic length scales into fracture theory may be essential to developing equations of motion for crack tips/fronts and to understanding dynamic instabilities. Whether or not these ideas are also relevant to the crack oscillations observed in rubber \cite{Deegan.02.rubber,Marder.JMPS.2006,Chen.11}, where the applied strains are extremely large and intersonic propagation velocities were reached, remains an open question.

\section{Three-dimensional tensile cracks: Micro-branching, front waves and corrugations}
\label{sec:3D}

In the previous section we discussed a way to understand the oscillatory instability based on a dynamic crack tip equation of motion that takes into account the length scale proposed by the weakly nonlinear theory of fracture mechanics. It was shown that a crack propagating in a 2D medium will lose its stability at extremely high velocities ($\sim\!0.9c_s$) to infinitesimal out-of-plane perturbations. We are still faced, however, with the challenge of explaining the micro-branching instability.
Micro-branching, on the face of it, appears to be a much more difficult problem. Once excited, this instability appears to be intrinsically 3D, as Fig.~\ref{instabilities2} suggests. It had been discovered in earlier work \cite{Livne.05} that the micro-branching threshold, $v_{mb}$, could vary above a {\em minimum} value of about $v_{mb}\!\simeq\!0.35c_R$. The observed values of $v_{mb}$ were shown to statistically depend on the value of the acceleration prior to the instability initiation. In addition, the transition to micro-branching was observed to be hysteretic in $v$.

These observations suggested that the micro-branching instability involves an activated process that could be triggered by noise above the minimum value of $v_{mb}$. If the acceleration is low, micro-branching is always observed near the minimal value of $v_{mb}$, as eventually the crack front will encounter a noise source that is sufficiently large to trigger the instability. In very thin gel samples, where ``sources of noise'' along the $z$ direction are geometrically reduced, and at high crack accelerations, micro-branching was seen to occur in a relatively large range of velocities, $0.35c_s\!<\!v_{mb}\!<\!0.9c_s$. This is the reason that the oscillatory instability could be, at all, observed prior to the onset of micro-branching~\cite{Livne.07,Goldman.12}.
\begin{figure}
\includegraphics[width=0.65\textwidth]{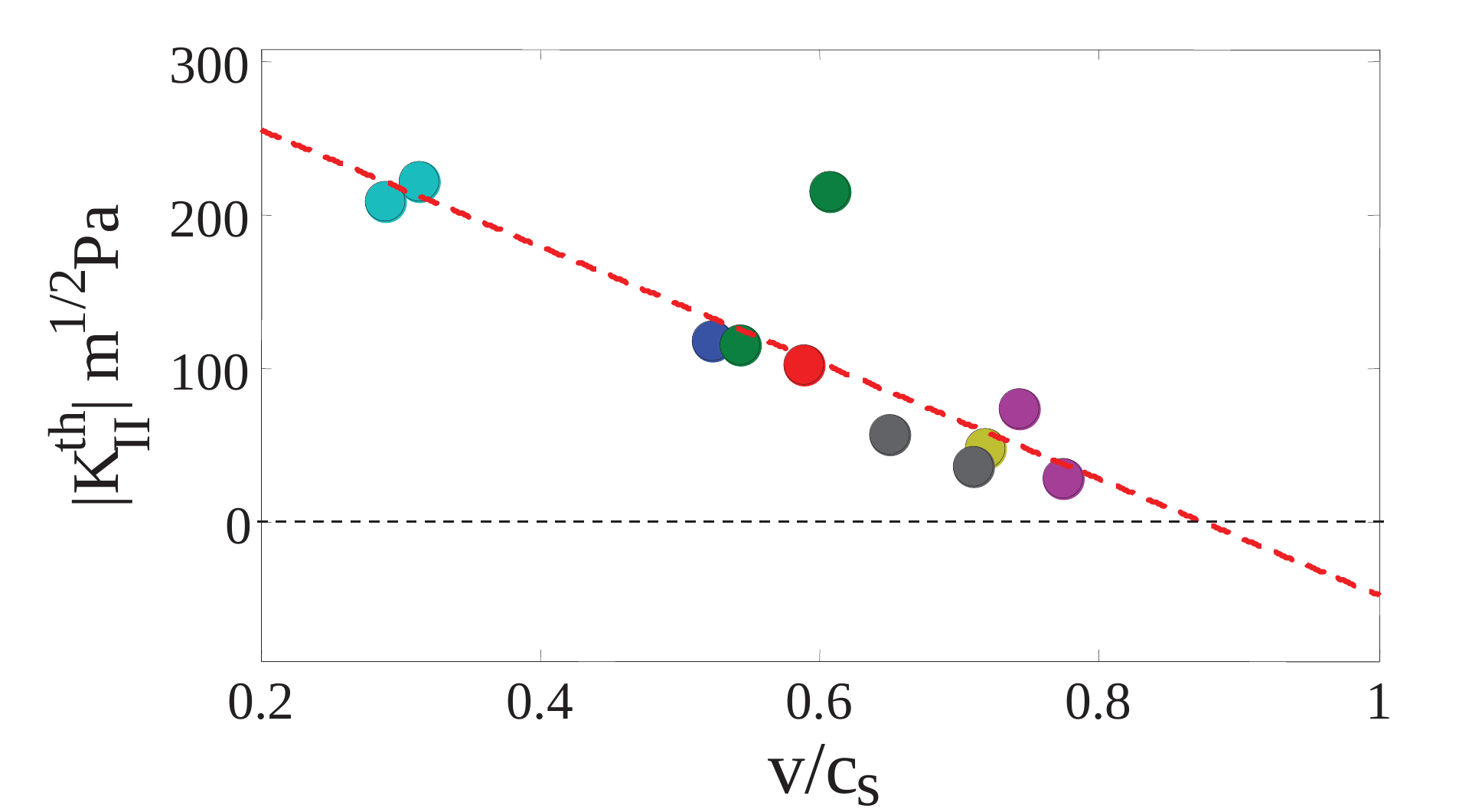}
\caption{Phase diagram of the micro-branching instability. The micro-branching instability is triggered when a small $K_{II}$  component is present. For each propagation velocity $v$, there exists a threshold value,  $ K^{th}_{II}(v)$.   When $ K_{II} > K^{th}_{II}(v)$  micro-branching is observed.  Shown are values $|K^{th}_{II}/K_I|$,which are the values obtained for each $v$ immediately prior to micro-branching. The critical boundary linearly extrapolates (dashed line) to zero at $v\!\simeq\!0.87 c_s$, the value of the measured threshold velocity for the onset of the oscillatory instability \cite{Goldman.12}. Colors denote different experiments. Adapted from~\cite{Goldman.15}.}
\label{threshold}
\end{figure}

Very recent experiments, however, suggest that the micro-branching instability may be closely related to the oscillatory instability~\cite{Goldman.15}. These experiments directly measured the 2D strain fields surrounding the tips of rapidly moving cracks in thin gels, as these cracks approached the instability over a wide range of $v_{mb}$ values. Prior to the onset of micro-branching, the cracks were ``simple'' so that the media could be considered as 2D. It was discovered that the strain fields of all of these simple cracks prior to the onset of the instability contained a fluctuating shear component that could be characterized by the textbook $K_{II}$ fields~\cite{Freund.90}. Examining the value of $K_{II}$ immediately prior to the instability onset, these studies showed that there exists a velocity-dependent threshold value, $K_{II}^{th}(v)$, for the onset of instability.

$K_{II}^{th}$ as a function of $v$ is presented in Fig.~\ref{threshold}. The data, when extrapolated to $K_{II}^{th}(v)\!\rightarrow\!0$, yield a zero threshold propagation velocity of $0.9c_s$, which is precisely the critical velocity for the onset of the oscillatory instability. Guided by local symmetry arguments \cite{Goldstein.Salganik}, we would expect any finite $K_{II}$ value to cause a crack to curve away from the initial symmetry axis imposed by the loading.
The data therefore suggest that two distinct ``phases'' can coexist in a hysteretic region $0.3c_s\!<\!v\!<\!0.9c_s$, where both straight cracks and cracks that locally curve out of the symmetry plane can be simultaneously stable. Local crack curvature will only occur upon a sufficiently large (finite) local shear perturbation, as shown in Fig.~\ref{threshold}.  Moreover, in this bi-stable region, both  straight and curved cracks can coexist within different sections along the crack front.  This bi-stability provides an explanation for the localized nature of the micro-branches in the $z$-direction. The bi-stability that enables this localization disappears at the critical velocity $0.9c_s$, where $|K_{II}|^{th}\!=\!0$. At this point the entire crack front must coherently curve out of plane and the oscillatory instability ensues.

Insofar as propagation along the symmetry axis is concerned, it is clear that straight sections of the crack front will outrun the curved sections. As a result, we expect that the crack front will eventually revert to a single fracture plane; pinching off the curved section. This ``frustrated curved section'' is what we identify as a micro-branch beneath the fracture surface. Such a  pinch-off  scenario  been recently observed experimentally by direct visualization of crack fronts in the $xz$ plane as micro-branching takes place~\cite{Kolvin.2015}. The micro-branch locally delays the crack front causing local crack front curvature within the $xz$ (fracture) plane. This initial curvature was seen to continually deepen, eventually developing into a cusp-like shape. Cusp formation within the fracture plane was coincident with the ``death'' of the micro-branch.

Micro-branching is, therefore, an intrinsically 3D instability that is intimately related to the crack front dynamics, and hence defies any attempt to understand it in a purely 2D framework. There are several other basic phenomena that necessitate a 3D framework of fracture dynamics. In the quasi-static regime, simulations coupled with analytic work  \cite{Lazarus.2001,Lin.2010,Lazarus.2008,Gao.1989,Xu.1994}   has shed light on a crack front instability under mixed-mode I+III conditions.Recently, helical perturbations have been shown, in 3D simulations,  to develop into segmented crack fronts \cite{Pons.10,Leblond.11,Leblond.2015}.Related segmented crack front structures have been observed in mode I~\cite{Baumberger.08,tanaka.98} and mixed-mode I+III fracture of of both ``standard" \cite{Smekal.1953,Sommer.1969,Knauss.1970,Cooke.96,Hull.97,Pham.2014} and  soft materials~\cite{Ronsin.14}.

In the dynamic regime, crack fronts were shown to support propagative modes along them. The most well-established of those are in-plane crack front waves, which were predicted through perturbative mode I numerical \cite{Morrissey1998,Morrissey2000} and analytic \cite{Ramanathan.97} calculations in the late 1990's. In this case, the crack front is assumed to remain within its tensile symmetry plane, but to experience in-plane shape perturbations. In the framework of LEFM and a linear perturbation theory \cite{Ramanathan.97,Morrissey1998,Morrissey2000}, perturbations to the crack front were shown to propagate along the front without attenuation (as long as $\Gamma$ is independent of $v$) at nearly $c_R$, for any $v$.

Subsequent experiments on various amorphous brittle materials \cite{Sharon.01,Sharon.02,Sagy.01,Sagy.04} indeed revealed the existence of localized modes that propagated along the crack front at nearly $c_R$, when triggered by an interaction with localized material inhomogeneities (asperities) or by spontaneously generated micro-branches. The experimentally observed propagative modes were different from the theoretically predicted in-plane front waves in two important aspects: (i) They featured an out-of-plane component, which left marks on the fracture surfaces and, in fact, enabled post-mortem measurements of crack front waves (in-plane perturbations do not leave any post-mortem fractographic signature) (ii) They were clearly nonlinear, exhibiting soliton-like properties \cite{Sharon.01}.  While  similar fracture surface markings can be generated by the interaction of shear waves with moving crack fronts \cite{Bonamy.2003,Sharon.2004,Ravichandar.2004}, recent measurements in both brittle gels \cite{Livne.05,Kolvin.2015} and the rapid fracture of rock \cite{Sagy.04,Sagy.06} have  strengthened the evidence for the existence of front waves in dynamic fracture.

Treating out-of-plane front perturbations is more difficult than in-plane ones because the front interacts with out-of-plane perturbations which are locked in the front's wake and because energy balance alone is insufficient to determine the front dynamics (information about the direction of propagation is essential as well, see discussion above on crack front equations of motion). Only very recently, based on the Willis-Movchan 3D LEFM linear perturbation formalism \cite{Willis.97} and the Principle of Local Symmetry, has the out-of-plane stability of crack fronts been considered~\cite{Willis.2012,Adda-Bedia.13}. In particular, a minimal scenario in which linearly unstable out-of-plane crack front corrugations might emerge (above a critical front propagation velocity), has been discussed. Within this scenario, a critical velocity has been calculated as a function of Poisson's ratio and corrugations have been shown to propagate along the crack front at nearly the Rayleigh wave-speed~\cite{Adda-Bedia.13}.

To linear order in the deviation from a straight crack front, in- and out-of-plane perturbations are decoupled. Taken together with the nonlinear nature of the experimentally observed front waves, nonlinear coupling between in- and out-of-plane front dynamics appears to be an essential next step. Due to the technical difficulties involved, numerical approaches such as phase-field models may be useful in achieving this \cite{Pons.10}. In addition, the effect of going beyond the Principle of Local Symmetry, e.g. in the spirit of \cite{Bouchbinder.09b}, should be systematically explored. Possible connections to the much studied problem of fracture surfaces roughness \cite{bonamy.11}, where dynamic effects are traditionally neglected~\cite{bouchaud.02}, should also be investigated.

\section{The dynamics of frictional interfaces: On the relation between friction and fracture, and more}
\label{sec:friction}
In this chapter we describe recent work that suggests that our understanding of many aspects of ``common-place" friction may need to be re-examined. This work suggests that dynamic fracture processes play a major role in frictional motion -- even when these processes appear to be extremely slow. The key to understanding how dynamic fracture comes into play is to consider the spatially extended interface that separates two bodies in frictional contact. In dry friction, this interface is composed of a large ensemble of rough and interlocking discrete contacts, whose total area of contact is orders of magnitude smaller than the nominal contact area of the contacting bodies \cite{Bowden.01,Dieterich.96}. In order for frictional motion to initiate, these contacts must be ``broken" or rupture. As we know, fracture of a spatially extended entity is not simultaneous, but is generally mediated by crack-like fronts. Below, we describe observations of such rupture fronts, which will be shown to be closely related to classic dynamic shear (mode II) cracks. It is intuitively obvious that these rapid rupture modes are also closely related to earthquakes, which mediate the relative motion of frictionally coupled tectonic plates.

The failure of macroscopic frictional interfaces has for centuries been thought to be governed solely by ``friction laws'', with the classic example of Amontons-Coulomb law of friction $F_s \!=\!\mu_{s,d} F_n$, where $F_s$ is the friction (shear) force and $F_n$ is the normal force. The proportionality constant is either the static friction coefficient $\mu_s$ or the dynamic one $\mu_d$, both are considered to be characteristic properties of the materials forming the frictional interface. This class of friction laws has been generalized to spatially varying and more physically faithful constitutive laws~\cite{Ruina.83,Dieterich.79,Marone.98,Scholz.98,Baumberger.06}. The latter, termed rate-and-state friction laws, incorporate the effects of additional dynamic variables such as the local sliding velocity (``rate'') and the local average contact lifetime (``state'') on friction.
\begin{figure}
\includegraphics[width=0.65\textwidth]{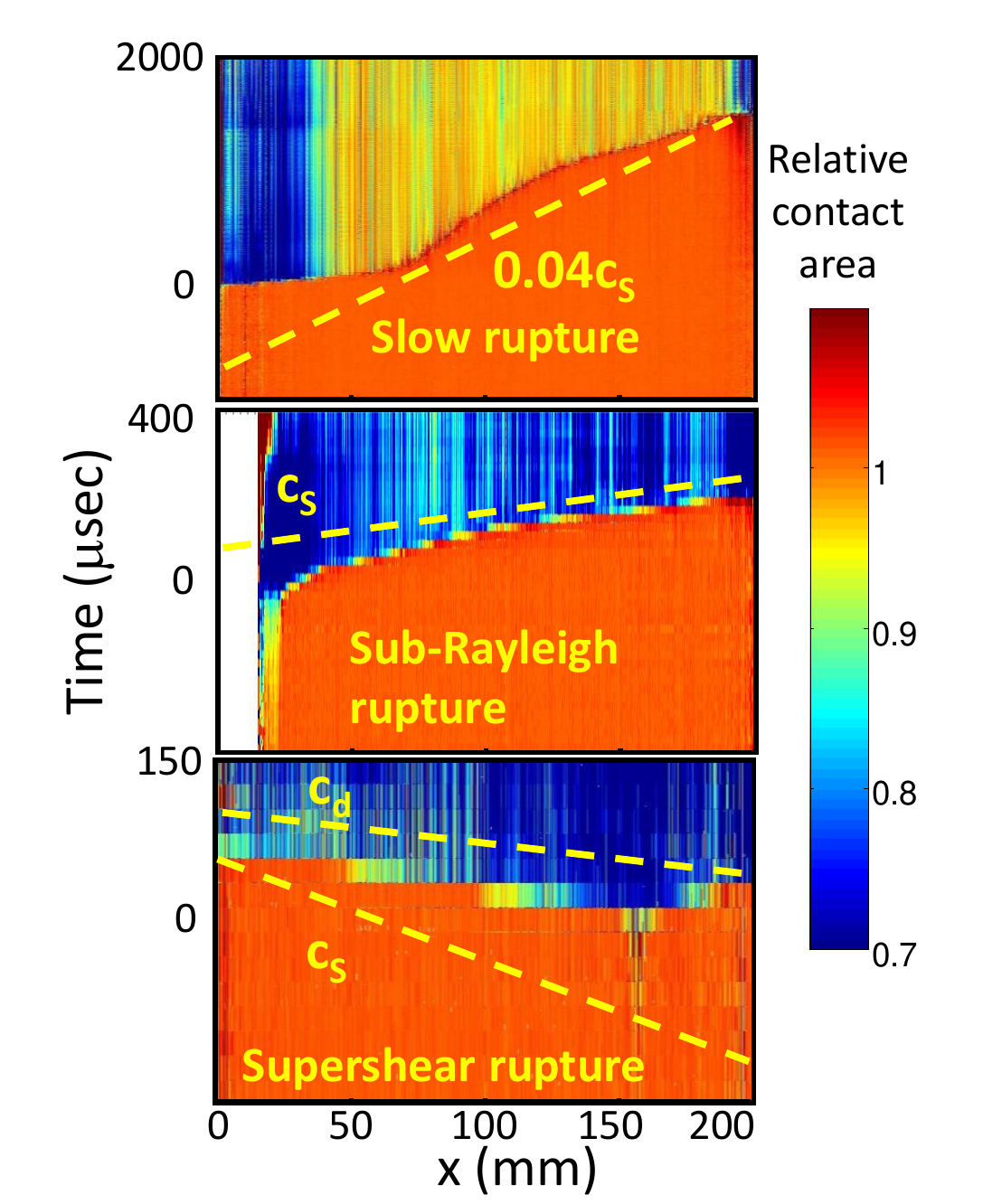}
\caption{Different types of rupture fronts in friction. Measurements of the real contact area reveal three qualitatively different types of fronts that ``fracture" the contacts defining the interface that separates two sliding bodies. Shown is the space-time variation of the real area of contact along a 200mm long frictional interface. The sharp reduction in the contact area occurs at the tip of the propagating fronts. Shown are typical examples of  (top) slow (center) fast (sub-Rayleigh) and (bottom) supershear ($v>c_s$) fronts. (adapted from \cite{Bendavid.10}).}
\label{fig:rupture fronts}
\end{figure}

A new generation of experiments over the past decade has revealed that rapid rupture fronts drive frictional motion. In other words, the transition from ``static'' to ``dynamic'' friction has been shown to be mediated by different types of crack-like fronts. These fronts are the vehicles that trigger the onset of macroscopic frictional motion, bridging the gap between the microscopic interactions that define local frictional resistance and the resulting macroscopic motion imbued in the slip of large bodies \cite{Das.03,Scholz.02,Ben-Zion2001,benzion.08}. As shown in Fig. \ref{fig:rupture fronts}, these new experiments have revealed various rupture modes that are loosely classified as follows: slow ruptures, which propagate far below the material wave speeds~\cite{Rubinstein.04,Latour.13,Bendavid.10,Rubinstein.11}, ``sub-Rayleigh'' ruptures \cite{Rubinstein.04,Latour.13,Bendavid.10,Rosakis.04} that propagate up to the Rayleigh wave-speed $c_R$, and ``super-shear'' rupture modes that surpass the shear wave-speed $c_s$ \cite{Rubinstein.04,Bendavid.10,Rosakis.04,Scholz.72,Rosakis.99,Passelegue.13}. It is important to note that analogs of all of these rupture modes have been documented in natural earthquakes \cite{Ben-Zion2001,benzion.08,Beroza.11,Bhat.07,Das.06,Bouchon.03}.
\begin{figure}[here]
\includegraphics[width=0.65\textwidth]{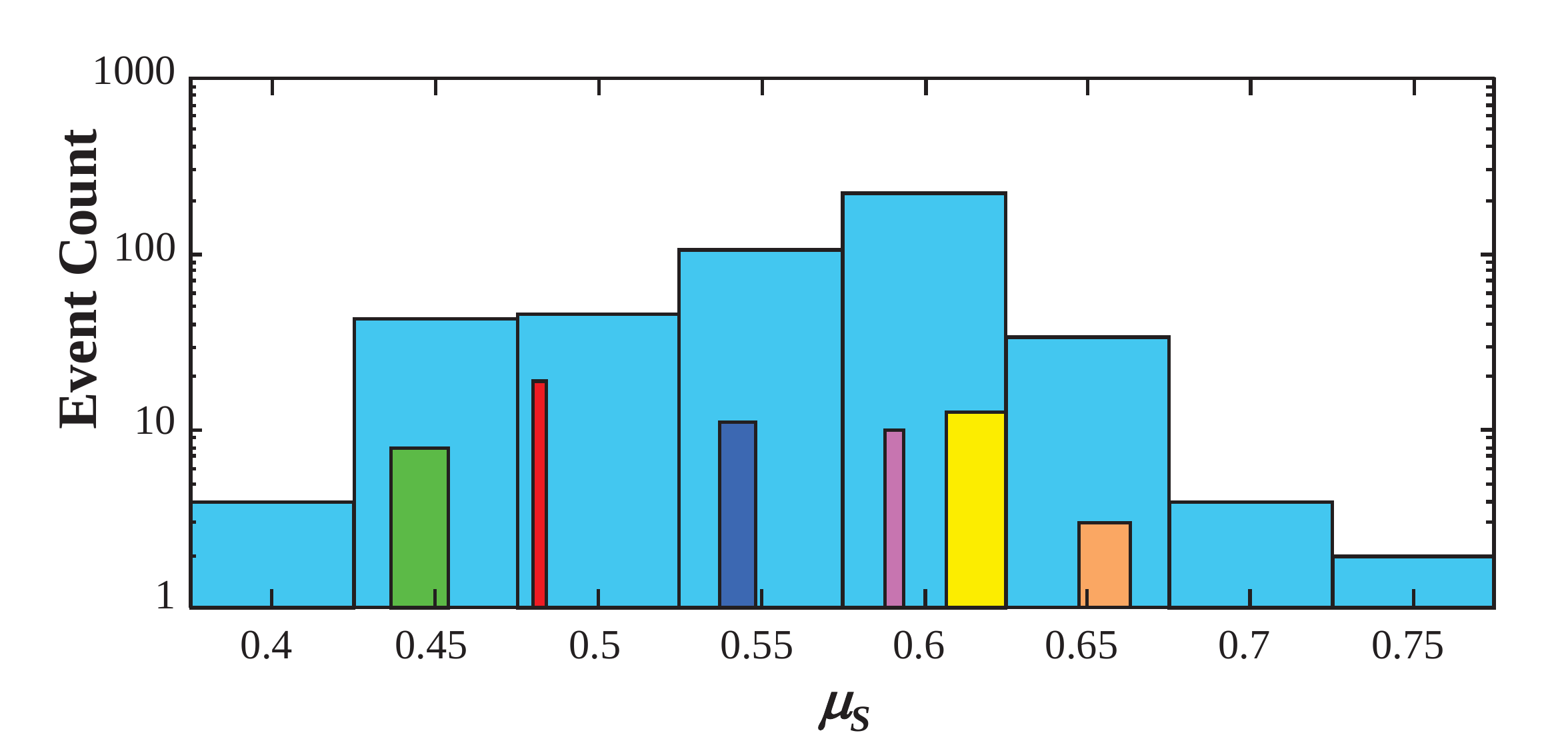}
\caption{Count of events with varying $\mu_s$, from experiments conducted under widely varying loading conditions. These range from pure edge loading to near uniform application of the shear force, $F_s$, to the system. Note that this plot is {\em not} a histogram and solely reflects the number of experiments performed whose loading conditions yielded $\mu_s$ for each range. The color bars (bar width = two standard deviations) in the plot depict sub-populations of events from single experiments, each with different (specific) loading conditions. Adapted from~\cite{Bendavid.11}.} \label{figmu}
\end{figure}

Are the rupture dynamics of an interface related to the values of the global ``friction coefficients" that characterise the interface strength? Recent experiments using PMMA~\cite{Bendavid.11} and granite~\cite{Passelegue.13} have measured the static friction coefficient, $\mu_s$, i.e. the ratio between the applied shear and normal forces at the onset macroscopic sliding. These studies found that sliding can be initiated over a wide range of static friction values, indicating that $\mu_s$ is certainly not solely a characteristic material property but is, instead, dependent on how a given system is loaded \cite{Bendavid.11}. Study of  the rupture dynamics leading to the broad range of friction coefficients, in fact, revealed that the loading conditions of the systems and the subsequent rupture dynamics were closely correlated to the value of $\mu_s$~\cite{Bendavid.11}. These experiments found that $\mu_s$ was as well defined as the loading conditions. As long as the way in which forces were applied to two blocks in frictional contact was kept constant, $\mu_s$  was extremely well-defined. Once the loading configuration was changed, the measured value of $\mu_s$ for the {\em same} two blocks could be varied significantly and controllably. This surprising result has been confirmed in recent theoretical and numerical studies \cite{Tromborg.11,Kammer.12,Nakano.10,Nakano.14}.

In addition to analytical work, recent numerical work~\cite{Tromborg.11,Kammer.12,Kammer.13,Kammer.15} together with simple models describing the contact dynamics of ensembles of discrete contacts have been successful in describing many aspects of frictional cracks \cite{Rubinstein.11,urbakh.09,Urbakh.11,Urbakh.12,Tromborg.14}. These results have provided important new insights into the influence of both the discrete nature of the contacts that compose the frictional interface and the influence of different ``friction laws" that couple the elastic medium to the interface.

How close to ``ordinary" fracture might the rupture processes shown in Fig. \ref{fig:rupture fronts} be? Let us consider two elastic bodies in frictional contact, subjected to remotely applied shear stresses. While in an intact homogeneous system the Principle of Local Symmetry would predict that any crack should rotate so as to annihilate any shear (mode II) component near its tip \cite{Goldstein.74}, the presence of a weak friction plane (compared to an intact bulk) allows rupture fronts to propagate under predominantly shear conditions along the weak plane. Are these rupture fronts related to classical mode II cracks? The surfaces left behind classical mode II cracks are assumed to be traction-free. While the propagation of rupture fronts along frictional interfaces is definitely accompanied by stress reduction (``stress drop''), it is clear that the frictional stress behind the propagating front is finite, and hence the physical situation is evidently different from the traction-free boundary conditions assumed in classical mode II crack solutions.

The two problems (i.e. of frictional rupture fronts and mode II cracks), though, are closely related when the following two conditions are met: (i) The residual stress $\sigma_r$ left behind the rupture front is constant, a situation commonly assumed in the context of earthquake dynamics \cite{Das.03,Scholz.02,Fruend.79,Burridge.79,Gabriel.2013} (ii) The width of the front (the ``cohesive zone''), i.e. the length scale over which the residual stress $\sigma_r$ is reached when the front passes a point at the interface, is small compared to other length scales in the system. In particular, under these conditions the singular mode II fields of LEFM remain valid (in this case the singular $K_{II}$ field mediates the transition from the initial shear stress to $\sigma_r$)~\cite{Palmer.73,Rice.80}.

These predictions have been directly tested in very recent experiments on PMMA blocks, where the real-time strain tensor at points adjacent to the interface was measured during rupture propagation~\cite{Svetlizky.14}. In parallel, the instantaneous rupture tip location and velocity were measured. These techniques, which were used successfully to describe the strain fields in mode I fracture \cite{Ravichandar.1983}, provided a first-time quantitative description of the strain fields surrounding the tips of these propagating rupture fronts. Measurements were performed for rupture fronts over a broad range of propagation velocities $v$, $0.04c_R\!<\!v\!<\!0.97c_R$.

\begin{figure}[here]
\includegraphics[width=.7\textwidth]{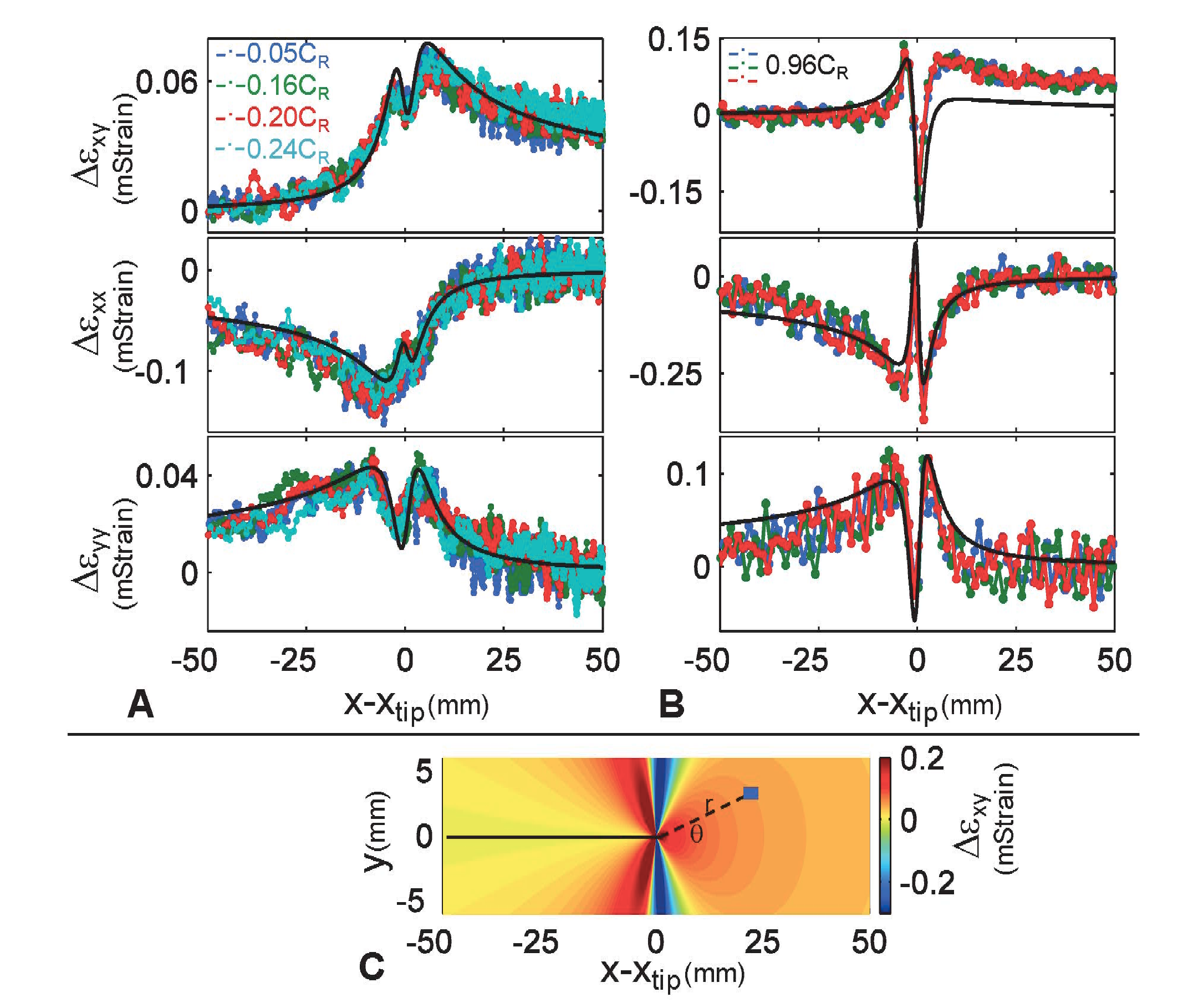}
\caption{(A) Measurements of the strain tensor, $\varepsilon_{ij}$, at a fixed spatial location above the frictional interface (see panel C for details) during the passage of a rupture front. The initial strain was subtracted from $\varepsilon_{xx}$ and $\varepsilon_{yy}$ and residual strain from $\varepsilon_{xy}$. Colors represent different, relatively small, front velocities ($0.05c_R\!<\!v\!<\!0.24c_R$). Corresponding LEFM predictions for plain strain boundary conditions are plotted in black ($v\!=\!0.2c_R$). The fracture energy $\Gamma\!\simeq\!1.1$J/m$^2$ is the sole free parameter. (B) $\varepsilon_{ij}$ for $v\!\simeq\!0.96c_R$. The LEFM solution (black), plotted as in panel A, still describes the larger amplitudes and strong oscillations of rapidly propagating ruptures with the {\em same} fracture energy as before. (C) Shear strain variations, $\varepsilon_{xy}$ , surrounding the rupture tip predicted by LEFM for $v\!=\!0.96c_R$. The blue square denotes the strain gage rosette measurement location relative to the approaching rupture tip. Note the strong angular dependence that drives the violent oscillations in panel B. Adapted from \cite{Svetlizky.14}.}
\label{strain fields}
\end{figure}

Typical results of these experiments are shown in Fig.~\ref{strain fields}, where a comparison of the different components of the strain tensor to the classic mode II singular solution for rupture fronts propagating at both slow velocities ($v\!<\!0.3c_R$) and velocities approaching the Rayleigh wave-speed, $c_R$, is presented. The surprisingly good agreement demonstrates that, at least in this material, the two conditions listed above are satisfied, and consequently the fields near the propagating rupture fronts are quantitatively described by the singular solutions predicted for shear (mode II) cracks in the framework of LEFM. It is interesting and important to note that the agreement to the LEFM singular fields was achieved with a velocity-independent fracture energy $\Gamma$, suggesting that the well-documented rate dependence of friction in PMMA~\cite{Baumberger.06} does not strongly influence the effective fracture energy.

The experiments described in Fig.~\ref{strain fields} suggest that propagating sub-Rayleigh ruptures can be identified with mode II cracks. Let us now consider the question of rupture nucleation. This question is of great importance both in general and in the particular context of earthquake nucleation (see, for example,~\cite{Benzion.2003,Ben-Zion1997,Lapusta2000,Rubin2005,Ampuero2008}. If nucleation takes place as a Griffith-like process analogous to conventional crack initiation~\cite{Lawn.93}, the nucleation length $L_c$ --- the critical size above which a slowly creeping patch becomes unstable --- is determined by the physical conditions near the rupture tip and depends on the pre-stress level. Alternative friction-law dependent nucleation scenarios, however, are possible. For example, rate-and-state friction models~\cite{Ben-Zion2001,benzion.08,Ben-Zion1997,Lapusta2000,Rubin2005,Ampuero2008} associate $L_c$ with a spatiotemporal friction instability of quasi-statically extending ``creep patches" that is independent of the pre-stress level~\cite{Lapusta2000,Barsinai.13}. It is currently unclear which nucleation scenario, if any, is more physically realistic and, if so, under what conditions.

 What is the origin and nature of the experimentally observed ``slow fronts'' \cite{Rubinstein.04,Bendavid.10}? Recent theoretical work has suggested that they may be a direct result of the qualitative form of the friction law at the interface; the frictional constitutive law can also affect the existence and nature of propagating rupture fronts. To see this, consider a frictional interface separating two infinitely long elastic blocks of height $H$. When the interface is described by a conventional velocity-weakening friction law, e.g. a friction law in which the frictional resistance decreases from $\mu_s$ to $\mu_d$ with increasing slip velocity $\Delta\dot{u}$, steady-state propagating rupture front solutions can be found under some fixed shear displacement loading conditions. These solutions are analogous to those found for tensile cracks in a strip~\cite{Freund.90}, where the only difference is the role played by the residual stress $\sigma_r$. On the other hand, no steady-state solutions can be found under shear stress controlled conditions; as in the analogous Mode I problem of constant stress, frictional cracks should continually accelerate to $c_R$ (as seen in Fig. \ref{equmotion}).

\begin{figure}
\includegraphics[width=0.55\textwidth]{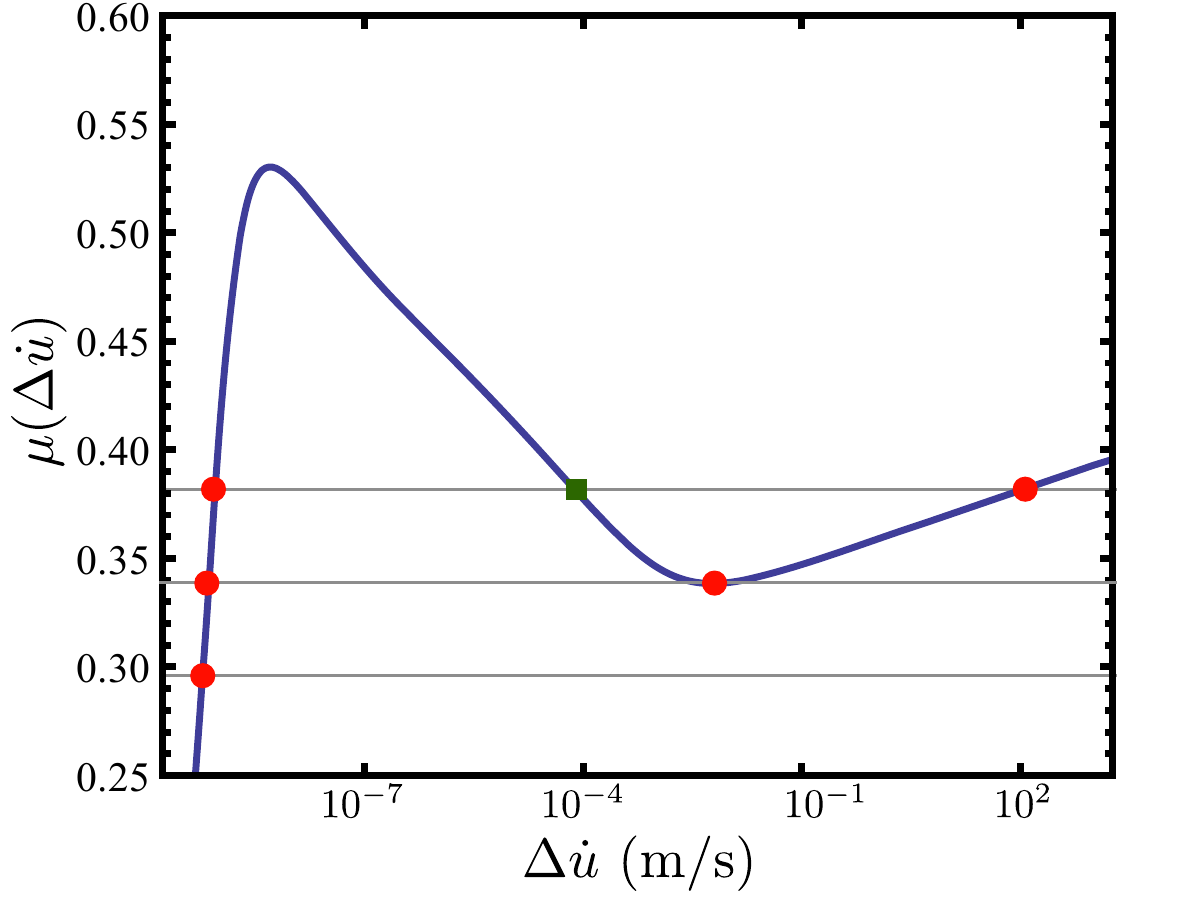}
\caption{A steady sliding friction curve $\mu(\Delta\dot{u})$, where $\Delta\dot{u}$ is the sliding velocity, which is typical of many materials~\cite{Barsinai.14}. The curve exhibits a non-monotonic behavior with a minimum after a small sliding velocity maximum. The three horizontal lines correspond to different levels of applied shear stress. A stable homogeneous solution is marked by a red circle and an unstable one by a green square.}
\label{fig:nonmonotonic}
\end{figure}

What happens for a non-monotonic friction law? Such a friction law, where the steady-sliding frictional resistance is a non-monotonic function of the sliding velocity, with a maximum at very small velocities and a higher velocity minimum, is depicted in Fig.~\ref{fig:nonmonotonic}. This form is representative of a broad range of materials \cite{Barsinai.14}. Consider again two infinitely long elastic blocks, now separated by an interface described by a steady sliding frictional resistance as in Fig.~\ref{fig:nonmonotonic}. A steady state propagating front is an object that smoothly connects two stable solutions in a homogeneous system, once the system is perturbed (e.g. via a stress gradient) and one of these solutions loses stability to the other. In the context of Fig.~\ref{fig:nonmonotonic}, stable solutions correspond to velocity-strengthening branches (i.e. when $\mu$ increases with increasing $\Delta\dot{u}$) and unstable (accelerating) solutions correspond to velocity-weakening branches (i.e. when $\mu$ decreases with increasing $\Delta\dot{u}$).

When the applied shear stress corresponds to the lower horizontal line in Fig.~\ref{fig:nonmonotonic}, below the minimum of the friction curve, there is only one stable homogeneous solution (marked by a red circle) and consequently no propagating front solutions exist. The situation is qualitatively different when the applied shear stress corresponds to the second horizontal line, which touches the friction curve at its minimum. In this case, there are two homogeneous solutions and a propagating front solution connecting the two is possible. Such a stably propagating solution was shown to exist \cite{Urbakh.11,Barsinai.13,Barsinai.12}, where the characteristic slip velocity corresponds to the minimum of $\mu(\Delta\dot{u})$. The solution's propagation velocity could be significantly smaller than the elastic wave-speeds and independent of them~\cite{Barsinai.13,Barsinai.12,Barsinai.15}. Further increase of the applied shear stress (e.g. the upper horizontal line) produces a continuous spectrum of propagating front solutions with increasing propagation velocities, which eventually saturates at the relevant wave-speed~\cite{Barsinai.12}. At stresses above the minimum of the friction curve, an unstable homogeneous solution also exists (green square in Fig.~\ref{fig:nonmonotonic}).

Such propagating front solutions do not exist for friction laws that monotonically decrease with increasing slip velocity. This is an example in which the properties of the friction law have qualitative implications for the existence of steady state rupture fronts. Transient rupture fronts, which emerge in more complicated situations, may be considered to be short-lived excitations of these steady state propagating rupture fronts~\cite{Urbakh.11,Barsinai.13,Barsinai.12}. The existence of these solutions may therefore provide a possible explanation for slow interface ruptures.  The possible roles of non-monotonic friction laws in naturally occurring slow earthquakes have been recently explored~\cite{Weeks1993,Shibazaki2003,Kato2003,Shibazaki01102007,Hawthorne2013a}. While other mechanisms for the emergence of slow rupture have also been proposed, they generally involve invoking other physical ingredients such as higher dimensionality, spatial variation of the friction law and stress heterogeneities~\cite{Yoshida2003,Liu.05,Rubin2008}, dilation and pore pressure effects~\cite{Rubin2008}, and a discrete description of asperities, including stochasticity of various local quantities~\cite{Tromborg.14}.

The frictional constitutive law can also give rise to crack-like objects which are not classical mode II cracks. For example, in the presence of viscous-friction (i.e. when the frictional resistance is linear in the sliding velocity, which may be relevant to both dry and lubricated frictional interfaces) the order of the displacement gradients singularity near the tip may differ from the common $\sqrt{r}$ singularity and depend on the friction law~\cite{Brener.02,Brener.05}. Furthermore, interesting physical properties such as a healing instability, a static threshold equivalent to the Griffith threshold and dissipation which is dominated by friction in a large interfacial region away from the tip region can emerge in certain situations \cite{Brener.02,Brener.05}. Such frictional shear cracks were shown to be related to the experimentally observed self-healing slip pulses along an interface between a gelatin gel and a glass~\cite{Baumberger.2002,Baumberger.2003,Ronsin.11}. These works established a clear connection between novel crack-like objects and frictional dynamics.

To this point we have described some recent insights obtained in the study of both sub-Rayleigh ($v\!<\!c_R$) and slow rupture fronts. We now briefly describe ruptures that propagate at velocities that are faster than $c_R$. These are commonly referred to as ``super-shear'' ruptures. Super-shear cracks have long been considered in the fracture literature \cite{Freund.90,Broberg.99,Andrews.76}, but until they were observed in the beautiful experiments by Rosakis and coworkers \cite{Rosakis.99}, they were often considered to be of purely theoretical value. Since this first observation in weakly bonded homogeneous blocks that were ballistically loaded, mode II super-shear ruptures have been observed numerically \cite{Gao.01,Das.14,needleman.08} as well as in a variety of other scenarios that include frictional interfaces and quasi-static loading~\cite{Rubinstein.04,Bendavid.10,Rosakis.04,Passelegue.13}.

These recent experiments have revealed that super-shear ruptures are not rare at all, but appear in many cases when the initial energy stored in the surrounding material is much higher than that needed to drive a running crack. When loaded ballistically, super-shear states are observed only for relatively high projectile velocities \cite{Rosakis.99}, whereas in frictional sliding super-shear ruptures are initiated when the initial shear stresses are very large (e.g. when the ``friction coefficient" is significantly greater than $\mu_s\!\simeq\!0.6$ in PMMA \cite{Bendavid.10} or granite \cite{Passelegue.13}). In addition to laboratory experiments and simulations, there is growing evidence that super-shear propagation can occur in earthquakes~\cite{Bouchon.03,Bouchon2001,Dunham.04,Dunham.12,Kanamori.14} which are often associated with extreme damage.

While their existence in no longer in doubt, a more precise characterization of super-shear ruptures is needed. It is not yet clear, for example, if the observed super-shear ruptures along frictional interfaces indeed correspond to ``textbook" super-shear solutions of mode II fracture (in analogy to the characterization performed in Fig. \ref{strain fields} for sub-Rayleigh ruptures).

The nucleation of super-shear ruptures is an important and still open question. While a number of scenarios for the nucleation of such modes have been suggested \cite{Rosakis.04,Broberg.99,Andrews.76,Dunham.03}, it is not yet entirely clear how the transition from sub-Rayleigh to super-shear ruptures comes about. A common theme among these different scenarios is that sufficient elastic energy becomes available in close proximity to a frictional interface. This has been shown to take place either via the interaction of the crack front with localized perturbations \cite{Dunham.03}, via radiation from an existing sub-Rayleigh front \cite{Rosakis.04,Broberg.99,Andrews.76,Madariaga.77}, or simply due to sufficient pre-existing strain energy surrounding the interface prior to nucleation \cite{Bendavid.10}. This last scenario has been suggested as a necessary one for mode I super-shear cracks in rubber-like materials~\cite{Marder.JMPS.2006,Chen.11,Gao.03} and may well be relevant for mode II cracks as well.

\section{Future directions}
\label{sec:future}

In the previous sections we highlighted a number of interesting topics in dynamic fracture that, in our opinion, have progressed significantly in recent years. Below we list a number of promising research directions that are related to each topic.

{\em Equation of motion for a crack.} The work reviewed in Sect.~\ref{sec:eom} has conclusively shown that the equation(s) of motion for simple cracks are correct, as long as a crack remains simple. The energy balance criterion works perfectly as long as the propagation direction of the crack is both prescribed and the relevant boundary conditions are properly accounted for. We do not have, as yet, a general equation for determining the path of dynamic cracks. One such equation was discussed in Sect.~\ref{sec:2Dinstabilities} and was applied to crack oscillations in 2D. Other interesting work has been performed in relation to dynamic path selection in crystalline materials \cite{Marder.EPL.04,Sherman.14}. Nevertheless, there is still much more to be done in this direction. Recent post-mortem analysis of fracture surface markings in PMMA~\cite{Bonamy.PNAS.13} has suggested that microscopic crack front motion takes place via the merging of voids ahead of the crack tip, at velocities significantly below the mean crack front velocity. Whereas these observations are perfectly compatible with LEFM, they raise an interesting question of size effects; how large do inhomogeneities have to be in order to invalidate the continuum description of crack front propagation? Whereas the interaction of crack fronts with single asperities has been explored in the LEFM framework \cite{Ramanathan.97,Morrissey1998,Morrissey2000}, it will be interesting to see how/if this framework generalizes to a large number/size of defects. Is there a scale where the crack front loses its identity as a coherent entity?

{\em Nonlinear elasticity and soft solids.} LEFM, as well as the weakly nonlinear theory, assume a ``base" state of zero background strain. This approximation works well for most ``standard" brittle materials, such as glass, brittle ceramics or brittle acrylics. Lately, new classes of materials are becoming available that join natural rubber as elastic, yet brittle, materials that can sustain large background strains prior to fracture. The fracture resistance of some of these synthetic materials, such as double-network hydrogels made of ionically and covalently cross-linked networks \cite{Gong.2003,Gong.2010,Sun.2012,Tetsuharu.2013,Vlassak.2014,Leibler.2014}, can be made to be enormous and new applications are expected to abound \cite{Marcellan.11,Marcellan_Nature.14}. In addition to such ``chemical gels'', there is also clear experimental evidence that elastic nonlinearities are important to fracture processes in ``physical gels"~\cite{Ronsin.14}, i.e. weakly cross-linked elastomers, that are important in biological applications. As all of these materials exhibit large strains prior to fracture, existing theoretical frameworks should be extended to large background strains. Very recent experiments have shown these effects to be important and the first extension of weakly nonlinear fracture mechanics to moderate background strains has been developed~\cite{Goldman.2015b}. This framework might also apply to the related topic of the strength of tough pressure-sensitive adhesives, such as adhesive tapes \cite{Ciccotti.13,Dalbe.14.1,Dalbe.14.2}.

{\em Instabilities in dynamic fracture.} The recent advances reviewed here may provide us with some tools needed to initiate a more complete understanding of dynamic instabilities. From these results, it appears that to fully understand these phenomena, the framework of fracture mechanics should be extended to systematically include 3D crack propagation. The near vicinity of the crack tip (e.g. process zone) is not simply a passive region that is needed to dissipate energy, but, as we saw in Sect.~\ref{sec:2Dinstabilities}, the process zone can actively feed back to the elastic fields driving the crack to produce unstable behavior. The 3D form of micro-branches implies that these effects entail a finite-size perturbation to the crack front. Initial studies of crack front waves considered only linear perturbations to an LEFM-described front. Further development of  a nonlinear description of crack front behavior may provide a way to better explore crack front interactions with either active noise sources (e.g. micro-branches) or quenched noise, such as material inhomogeneities. Other promising directions to describe 3D effects include the use of phase-field models to provide an accessible way to couple the nonlinear elastic fields to dynamic models describing the crack tip. Some initial steps in this direction have been accomplished~\cite{Henry.13,Leblond.11,Karma_Nature.10}, but there are still a number of open questions about how to ensure the physical predictive power of these models.

{\em Fracture-based descriptions of friction.} Recent experiments have demonstrated that frictional motion is intimately related to dynamic fracture. A fundamental understanding of these important processes must, therefore, involve rapid fracture. While many related issues has been extensively discussed in the geophysical literature~\cite{Das.03,benzion.08}, careful examination of frictional ``fracture" in well-controlled laboratory experiments has only began to emerge. In Sect.~\ref{sec:friction}, a few aspects of this research direction have been mentioned. Different modes of earthquake-like rupture states are expected to abound in these systems; these range from sustained slow fracture modes to super-shear rupture.  The results described in  Fig. \ref{strain fields}, for example, relate to dry and rough frictional interfaces formed by {\em identical} materials. It would be interesting to make a similar comparison to analytical results that describe mode II super-shear cracks along homogeneous interfaces, whose singular properties are predicted to be quite different from those of standard cracks \cite{Broberg.99}. Entirely different classes of rupture, that include ``self-healing'' pulse-like as well as crack-like rupture states, are believed be generated within {\em bimaterial} interfaces; interfaces formed by two materials having different elastic properties \cite{benzion.08,Weertman.80,Ben-Zion.98,Rice.00a,Gerde.01,Perrin1995}. Recent experimental results \cite{Rosakis.06,Rosakis.10} have demonstrated the existence of a number of these interesting modes. Crack-like modes may also be excited within lubricated interfaces \cite{Brener.02,Brener.05}. Like super-shear states, these rupture modes are predicted to possess functional forms that are still singular, but with singularities that may be entirely different from the square root singularity generally associated with cracks.

In conclusion, in this perspectives piece we have presented a number of new and interesting developments in the field of dynamic fracture. The examples noted are all important, but are by no means all-inclusive. We have endeavored to show that dynamic fracture is not only important in itself, but is critical to our fundamental understanding of topics ranging from the strength of new ``super-tough" materials to the dynamics of earthquakes. We expect that these developments will continue into the next decades.

%
%

\begin{acknowledgements}
E. B. and J. F. acknowledge support from the James S. McDonnell Fund (Grant No. 220020221), E. B. acknowledges support from the Minerva Foundation with funding from the Federal German Ministry for Education and Research, the Harold Perlman Family Foundation and the William Z. and Eda Bess Novick Young Scientist Fund. J. F. acknowledges support from the European Research Council (Grant No. 267256), and the Israel Science Foundation (Grant 76/11).
\end{acknowledgements}


\end{document}